\documentclass[aps,pra,twocolumn,superscriptaddress]{revtex4-1}

\usepackage{graphicx,subfig}
\usepackage{hyperref,todonotes}
\usepackage{amsmath,mathrsfs,braket,bm,dsfont,amsfonts}

\begin{document}
\title{Robustness of Adaptive Quantum-Enhanced Phase Estimation}
\author{Pantita Palittapongarnpim}
\affiliation{%
	Institute for Quantum Science and Technology,
    University of Calgary, Alberta T2N 1N4, Canada}
\author{Barry C. Sanders}
\affiliation{%
	Institute for Quantum Science and Technology,
    University of Calgary, Alberta T2N 1N4, Canada}
\affiliation{%
	Program in Quantum Information Science,
    Canadian Institute for Advanced Research,
	Toronto, Ontario M5G 1M1, Canada}

\date{\today}

\begin{abstract}
As all physical adaptive quantum-enhanced metrology schemes operate under noisy conditions 
with only partially understood noise characteristics, 
so a practical control policy must be robust even for unknown noise.
We aim to devise a test to evaluate the robustness of AQEM policies and assess the resource used by the policies.
The robustness test is performed on adaptive phase estimation by simulating the scheme under four phase noise models corresponding to the normal-distribution noise, the random telegraph noise, the skew-normal-distribution noise, and the log-normal-distribution noise. 
The control policies are devised either by a reinforcement-learning algorithm in the same noise condition, albeit ignorant of its properties, or a Bayesian-based feedback method that assumes no noise.
Our robustness test and resource comparison can be used to determining the efficacy and selecting a suitable policy.
\end{abstract}
\maketitle

\section{Introduction}
\label{sec:introduction}
Quantum-enhanced metrology (QEM) employs a quantum state of~$N$ particles as a resource
to estimate an unknown parameter~$\phi$ with the goal of attaining imprecision
\begin{equation}
\label{eq:imprecision}
	\Delta\tilde{\phi}
    	\in O\left(N^{-\wp}\right),
\end{equation}
that asymptotically surpasses the standard quantum limit (SQL)
$\wp=1/2$~\cite{GLM04,GLM11}
but saturates at or below the Heisenberg limit (HL) $\wp=1$~\cite{BCM96,TA14}.
Our expression~(\ref{eq:imprecision})
differs from usual proportionality expressions in the literature,
e.g., $\Delta\tilde{\phi}\propto N^{-\wp}$
by explicitly recognizing the relevance of lower-order terms~\cite{TA14}
through the use of the big-O notation~\cite{VK14_ch1}.

QEM is vital for high-precision applications, such as gravitational wave detection~\cite{Hol79,CTD+80,Cav81}, atomic clocks~\cite{BIWH96,BS13}, and magnetometry~\cite{TA14,DLV+18} whose systems are operating at the limit of their power tolerance.
Some schemes consider ideal measurements that typically involve measuring multiple particles simultaneously~\cite{ZD14,MZC+16},
whereas adaptive QEM (AQEM) focuses on single-particle measurements augmented by feedback
such that the SQL is beat and the HL is approached~\cite{Wis95,AAS+02,BW00}.

AQEM Performance critically depends on policy choice~\cite{RZB18},
which can be obtained by optimizing a known mathematical model~\cite{RPH15,WK97,WK98}
or by using reinforcement-learning algorithms~\cite{HS10,LCPS13,PWZ+17}.
Whereas policies from these methods are resistant to known noise models~\cite{RPH15}, whether they are robust against unknown noise is yet unstudied but critical property of a QEM scheme as noise can destroy the entanglement advantage and restore the SQL~\cite{EMD11,DKG12}.
Our aim is to test the robustness of AQEM policies in the presence of noise with unknown properties. 

Our test focuses on adaptive interferometric phase estimation, 
whose policies have been devised using Bayesian techniques~\cite{BW00,BWB01} and by reinforcement learning~\cite{HS10,LCPS13,PWZ+17}. 
The Bayesian technique computes feedback based on a trusted, noiseless quantum model, 
whereas reinforcement learning devises the AQEM policies for feedback based on trial and error and is ignorant of the quantum-dynamical nature but employs heuristics to shrink the search space.
Here both methods are applied to adaptive phase estimation including phase noise, 
which could arise from path-length fluctuation in the interferometer~\cite{Bob87,SGS+14}.

Typically, noise is assumed to be normal as a result of the central limit theorem~\cite{Sev05_ch12}.
Periodicity of phase makes the normal distribution problematic unless the noise is small compared to $2\pi$ radians,
which we assume here;
technically,
we would use the wrapped-up normal distribution~\cite{Bre63}.
As our aim is to test robustness for unknown  noise, we consider three other noise distributions for our test: 
random telegraph~\cite{New68}, 
skew-normal~\cite{AC14}
and log-normal~\cite{LSA01} noise.
The random telegraph noise simulates a discrete noise process. 
Skew-normal and log-normal distributions
represent asymmetric noise,
which serve as distinct generalizations of the normal distribution.
Both distributions are used to simulate noise in detectors and electronics~\cite{Pou07,Jou11,RE14}.

For AQEM, we seek an efficient procedure that beats the SQL, and we choose the policy that requires the least resource to run.
We assess the policy-generating procedure according to the complexity of its time cost~\cite{VK14_ch5}, which is evaluated by the scaling in the number of operations with the number of particles $N$.
Here we consider two policy-design procedure, namely, a reinforcement-learning algorithm and a method based on Bayesian inference, resulting one policy designed by each method.
To determine which policy is superior, we compare the complexity in space and time cost~\cite{VK14_ch1}.
Thus, we are able to assess and compare the costs for 
generating policies and determine the best policy.

Through our analysis,
we find that both Bayesian-feedback and reinforcement-learning policies are robust in the face of unknown noise.
Specifically,
the Bayesian method yields AQEM that approaches HL and outperforms reinforcement-learning policies for most noise models. 
This performance superiority is due to the Bayesian method memorizing the measurement history through complete knowledge of the quantum state.
Storing the entire model in the computer yields better scaling
but leads to higher space and operational time costs compared to the reinforcement-learning policy. 

Disparity between the space costs suggest that the two approaches use different amounts of information to execute the feedback control, and a fair comparison should be made instead when they are using the same space cost. 
In principle, the reinforcement-learning policy can be generalized to attain the HL as long as the design cost, which scales in high-polynomial, is practical. 

\section{Background\label{sec:background}}
In this section, we present essential background knowledge for assessing robustness of adaptive quantum-enhanced phase estimation.
In this subsection, we cover four key notions as subsections.
In \S\ref{subsec:QEM} we discuss QEM including AQEM. Subsequently,
in~\S\ref{subsec:phase estimation},
we address a particular case of AQEM, 
namely, adaptive phase estimation. As noise is an important in practical AQEM, in \S\ref{subsec:noisemodels} we discuss noise models that we use to evaluate policy robustness.
Finally,
in \S\ref{subsec:regression},
we review regression analysis and model selection techniques that are required for analyzing whether the SQL has been surpassed.

\subsection{Quantum-enhanced metrology\label{subsec:QEM}}
In this subsection, we explain QEM strategies, mainly collective non-adaptive~\cite{GLM11} 
and adaptive measurements~\cite{PWS16b}, 
and lower imprecision bounds attained by using the classical and quantum resources.
Specifically, we focus on strategies that employ finite~$N$ input states.
Whereas the collective non-adaptive strategy is useful to calculate lower bounds for imprecision,
adaptive strategies offer the simpler alternative of individual-particle measurements that can achieve imprecision scaling close to these theoretical bounds.
Here we focus on single-parameter estimation; 
multi-parameters QEM~\cite{Hel69,HBDW13,YZF14}
would be the subject of future study.

\subsubsection{Non-adaptive and adaptive strategies}
Here we describe collective non-adaptive and adaptive QEM strategies, 
which are two of many types of QEM schemes.
Whereas we focus on these two techniques, there are others, 
such as the sequential technique~\cite{HBB+07,DM14} and the ancilla-assisted techniques~\cite{HKO05,SGM+18}, 
which we do not cover.

\paragraph{Collective strategy.}
A collective non-adaptive scheme utilizes~$N$ $d$-level
(typically, $d=2$ for standard two-level atoms or two-path interferometry)
particles 
prepared in a collective state
\begin{equation}
\label{eq:rhoHdN}
	\rho
    	\in\mathcal{S}\left(\mathscr{H}_d^{\otimes N}\right),
\end{equation}
which is the space of positive-definite, trace-class, self-adjoint linear operators acting on a tensor product of~$N$ copies of a $d$-dimensional Hilbert space~$\mathscr{H}_d$~\cite{Lan17}.
The system,
upon which metrology is performed,
is represented as a quantum channel
(completely-positive trace-preserving map)~$\mathcal{I}(\phi)$, acting on any~$\rho$,
with~$\phi$ the single-unknown parameter of the channel~\cite{Hol12}.
In the special case of an isolated system without noise, decoherence or loss, 
the channel is represented by a unitary transformation~\cite{YMK86}
\begin{equation}
\label{eq:IU}
	\mathcal{I}(\phi)\rho
    	=U(\phi)\rho U^\dagger(\phi)
\end{equation}
for~$U$ a unitary operator acting on~$\mathscr{H}_d^{\otimes N}$.

After the particles exit the system, they are measured and this measurement is described a positive-operator-valued measures~\cite{Hay2006,WM09}, 
which are positive semidefinite operators
\begin{equation}
\label{eq:X_m}
	\hat{X}_x:
    	\mathscr{H}_d^{\otimes N}
        	\to\mathscr{H}_d^{\otimes N},
	\sum\limits_x \hat{X}_x=\mathds{1},
\end{equation}
assuming the measurement outcomes $\{x\}$ is a finite set. 
This outcome is random with probability
\begin{equation}
\label{eq:Px}
P_x=\text{tr}\left(\hat{X}_x\mathcal{I}(\phi)\rho\right).
\end{equation}
The measurement~$\hat{X}_x$
is repeated multiple times to sample the distribution~(\ref{eq:Px}) sufficiently well to get good estimates,
and then $\phi$ is then inferred from these samples.

\paragraph{Bundle and individual-particle measurement.}
Instead of collective measurement,
we can consider measuring subsets of particles,
which we call bundles,
and, at the extreme limit,
which is of interest here,
the individual-measurement case of measuring a single particle at a time.
Mathematically,
we split the particles into~$M$ bundles of~$L$ particles where~$N=ML$~\cite{PWS16b}
so the Hilbert space can be expressed as
\begin{equation}
\label{eq:HdML}
	\left(\underbrace{\mathscr{H}_d^{\otimes L}}_\text{bundle}\right)^{\otimes M}.
\end{equation}
In this case, both~$\mathcal{I}(\phi)$ and~$\hat{X}_x$ act on~$\mathcal{H}_d^{\otimes L}$.
For localized measurements on each bundle,
the POVM is
\begin{equation}
\label{eq:POVMm}
	\bigotimes_{m=0}^{M-1}\hat{X}_{x_m}^{(m)},\;
    \hat{X}_{x_m}^{(m)}:\mathscr{H}_d^{\otimes L}
    	\to\mathscr{H}_d^{\otimes L}
\end{equation}
with outcomes from this tensor-product POVM being concatenations of~$M$ length~$L$ strings of $d$-dimensional digits,
\begin{equation}
\label{eq:xdML}
	\bm{x}_M
    	=x_0x_1\cdots x_{M-1}
    	\in\mathbb{N}_{d^L}^{\otimes M},
\end{equation}
where
\begin{equation}
\label{eq:xidM}
	x_m\in\mathbb{N}_{d^L}
    	:=\{0,1,2,\dots,d^L-1\}
\end{equation}
measured form the~$m^\text{th}$ bundle.

In one extreme case, each bundle contains only one particle,
which leads to $M=N$ and $L=1$. The string of outcomes becomes
\begin{equation}
\label{eq:xdN}
	\bm{x}_N
    	=x_0x_1\cdots x_{N-1}
        \in\mathbb{N}_d^{\otimes N}.
\end{equation}
The POVM is
\begin{equation}
\label{eq:Xmsingle}
	\bigotimes_{m=0}^{N-1}\hat{X}_{x_m}^{(m)},\;
    \hat{X}_{x_m}^{(m)}:
    	\mathscr{H}_d\to\mathscr{H}_d,
\end{equation}
which is a tensor product of~$N$ qudit POVMs.

For two-level particles, 
the state~(\ref{eq:rhoHdN})
is simplified to
\begin{equation}
\label{eq:rhoH2N}
	\rho
    	\in\mathcal{S}
        	\left(\mathscr{H}_2^{\otimes N}\right),
\end{equation}
and the POVM simplifies from~(\ref{eq:Xmsingle}) to
\begin{equation}
\label{eq:Xmsingle2}
	\bigotimes_{m=0}^{N-1}\hat{X}_{x_m}^{(m)},\;
    \hat{X}_{x_m}^{(m)}:
    	\mathscr{H}_2\to\mathscr{H}_2.
\end{equation}
The outcome~(\ref{eq:xdN})
is simplified to
\begin{equation}
\label{eq:x2}
	\bm{x}_N\in\left\{0,1\right\}^{\otimes N},
\end{equation}
which is an $N$-bit string.
Henceforth,
we restrict to the $d=2$ (two-level system),
$L=1$ (single-particle-per-bundle case)
for simplicity and without loss of generality.
\paragraph{Adaptive strategy.}
The adaptive strategy involves incorporating quantum feedback control~\cite{JN15}
such that the system operation depends on both the unknown parameter~$\phi$
and a control parameter~$\Phi_m$,
for some degree of freedom,
on the $m^\text{th}$ bundle.
We assume that incorporating a control
preserves the system acting as a channel
and thus write the channel acting on the $m^\text{th}$ bundle as
$\mathcal{I}\left(\phi;\Phi_m\right)$.
Measurement of the $m^\text{th}$ bundle
leads to an update of the control parameter to~$\Phi_{m+1}$ for the next bundle.

The control-parameter update is determined by a policy
\begin{equation}
\label{eq:policyupdate}
	\varrho:
    	\left(\bm{x}_m,\Phi_m\right)\mapsto\Phi_{m+1},\;
        \bm{x}_m=x_0x_1\cdots x_{m-1},
\end{equation}
which uses the string of outcomes~$\bm{x}_m$
and the most recent control parameter~$\Phi_m$
to obtain the next control parameter~$\Phi_{m+1}$.
This procedure continues
until reaching the final particle,
i.e., the $N^\text{th}$ particle,
at which point the estimate of the unknown parameter is
\begin{equation}
\label{eq:tildephiPhiN}
	\tilde{\phi}:=\Phi_N.
\end{equation}
Therefore, the adaptive strategy can be used for single-shot measurement; i.e., inferring~$\phi$ from one instance of the measurement procedure.

\subsubsection{Imprecision limits}
\label{subsubsec:imprecisionlimits}
Imprecision of the estimate~(\ref{eq:tildephiPhiN})
is denoted $\Delta\tilde{\phi}$~(\ref{eq:imprecision}).
Assuming the measurement is optimal
and that the quantum channel is noiseless~\cite{CTD+80,Sch14},
imprecision lower bounds are calculated for classical and quantum resources.
These lower bounds are the SQL and HL,
respectively.
In a noisy system,
which pertains in practice,
a QEM scheme is unlikely to saturate the bound. 
Despite the presence of noise,
the SQL is still the bound that must be surpassed to claim QEM,
which requires using quantum resources.
Here we review these limits 
as benchmarks for both the robustness test and to compare AQEM policies.

\paragraph{Standard quantum limit.}
The SQL is the imprecision lower bound if classical resources are used,
which means that the input state~(\ref{eq:rhoH2N})
is separable,
i.e., unentangled~\cite{KD13}.
The simplest case of such a separable state is a tensor product of~$N$ independent particles~\cite{GLM04},
\begin{equation}
\label{eq:untangled state}
	\rho=\rho_1^{\otimes N}.
\end{equation} 
Interacting this state to the quantum channel leads to the output state
\begin{equation}
\label{eq:indepedent output}
\left(\mathcal{I}(\phi)\rho_1\right)^{\otimes N}.
\end{equation}
Measuring this state~(\ref{eq:indepedent output})
according to the POVM~(\ref{eq:Xmsingle2})
leads to output governed by the probability distribution,
\begin{equation}
	P(x_m)
    	=\text{tr}\left(\hat{X}_{x_m}\mathcal{I}(\phi)\rho_1\right),
\end{equation}
which is independent and identically distributed (iid) for~$m\in\{0,1,\dots,M-1\}$.
Calculating the imprecision,
such as through the central limit theorem, leads to $\wp=1/2$ because of this iid condition~\cite{Mac13}.
The scaling also holds for the imprecision lower bound, which is calculated from~(\ref{eq:indepedent output})
using the Cram\'{e}r-Rao lower bound~\cite{Kay93}, and is irrespective of the quantum channel.

\paragraph{Heisenberg limit.}
If quantum resources are employed,
e.g.,
squeezing~\cite{Brau05} or entanglement~\cite{Woo98},
the SQL can be surpassed~\cite{Cav81,TA14}.
The lower bound to using the quantum resource can be computed from the quantum version of the Cram\'{e}r-Rao lower bound~\cite{Hel69}, which depends on the input state and the quantum channel~\cite{Jacobs2014_ch6}.
Therefore, unlike the SQL, the HL is specific to the QEM scheme~\cite{ZPK12}.
In the case of interferometric phase estimation, the HL is known to be $\wp=1$~\cite{BS84}, 
although this limit can only be attained through the use of optimal measurement.
As an optimal POVM could be infeasible,
adaptive phase-estimation schemes provides an attractive alternative to achieving close to this lower bound~\cite{WBB+09}.
	
\subsection{Adaptive phase estimation\label{subsec:phase estimation}}
Phase estimation underlies many QEM applications~\cite{Cav81,YMK86,GLM04}
and thus is widely used for devising quantum-enhanced techniques, including several AQEM schemes~\cite{WBB+09,HS10,RPH15}.
Here we explain the interferometric adaptive phase-estimation scheme controlled
by Bayesian feedback~\cite{BW00,BWB01} or reinforcement-learning policies~\cite{HS11b,LCPS13,PWZ+17}, which we compare in terms of robustness and resource consumption for control.

\subsubsection{Adaptive interferometric phase-estimation procedure}
One method of estimating phase is to use an interferometer, which infers phase shifts from the interference between two or more modes~\cite{Arm66}. 
In particular, we use an adaptive phase-estimation scheme based on a Mach-Zehnder interferometer,
which has two modes and therefore we are looking at the case of~$d=2$ representing the modes.
The mathematics of Mach-Zehnder interferometry applies to other forms of SU(2) interferometry, 
such as Ramsey, Sagnac and Michelson interferometry~\cite{YMK86,HS96,GLM04}.
In this subsubsection,
we present the input state,
adaptive channel,
detection, feedback, inference and imprecision.
\paragraph{Input state.}
For non-adaptive quantum interferometry with collective measurement,
the unitary interferometric transformation is in the Lie group SU$(2)$
with irrep (Casimir-invariant label)
$j=N/2$.
For adaptive quantum interferometry or individual measurements,
the interferometric unitary transformation is SU$(2^N)$ for~$N$ particles and two paths.
However, the two descriptions converge if the input state is permutationally symmetric;
technically,
Schur-Weyl duality dictates that the applicable transformation is SU(2) with irrep $j=N/2$~\cite{HS11a}.

Notationally,
modes are labelled by
\begin{equation}
	\varepsilon_m\in\{0,1\},
\end{equation}
which conveys which of the two paths,
such as input or output port or intra-interferometric path,
pertains.
Thus, the state~$\ket{\epsilon_m}$
refers to $m^\text{th}$ photon being in path~$\epsilon_m$.
The multiphoton basis is the tensor-product state
\begin{equation}
	\ket{\bm{\epsilon}_N}
    	=\bigotimes_{m=0}^{N-1}
        	\ket{\epsilon_m}.
\end{equation}
For $\operatorname{ham}\bm{\epsilon}$
the Hamming weight,
i.e., sum of bits, of~$\bm{\epsilon}$,
the permutationally-symmetric basis is
\begin{equation}
\label{eq:symmetricbasis}
	\ket{n,N_a-n}
		=\binom{N_a}{n}^{-1/2}  	\sum_{\operatorname{ham}\bm{\epsilon}_{N_a}}
        	\ket{\bm{\epsilon}_{N_a}}
\end{equation}
for~$N_a$
the total number of particles in mode~$a$.

The sine state serves as a loss-tolerant symmetric state that minimizes phase-estimation imprecision~\cite{SP90,WK97,WK98}, 
and is expressed as~\cite{BW00,BWB01}
\begin{align}
\label{eq:sinestate}
\ket{\psi}_N
=&
\left(\frac{N}{2}+1\right)^{-1/2}
\sum\limits_{n,k=0}^{N}\sin\left(\frac{k+1}{N+2}\pi\right)\text{e}^{\text{i}\pi (k-n)/2}
\nonumber\\&
\times
d_{n-N/2,k-N/2}^{N/2}\left(\frac{\pi}{2}\right)
\ket{n,N-n},
\end{align}	
for $d_{m,m'}^{j}\left(\beta\right)$
the Wigner-$d$ function~\cite{Mis14}.
This state also has the advantage of being robust against photon loss~\cite{HS11a}, 
which is a desirable property for practical QEM and so is used in the adaptive phase estimation procedures.
\paragraph{Adaptive channel.}
The particles in the sine state~(\ref{eq:sinestate})
are divided into single-particles bundles
($L=1$ case), 
each of which passes through the Mach-Zehnder interferometer.
For a noiseless interferometry, the quantum channel for one photon is
\begin{equation} 
	U_1(\phi;\Phi_m)
    	=\exp(\text{i}(\phi-\Phi_m)
        	\hat{\sigma}_y),\;
    \phi,\Phi_m\in[0,2\pi)
\label{eq:MZI unitary}
\end{equation} 
for~$\hat{\sigma}_y$ a Pauli matrix~\cite{SM95}.
Therefore,
the channel is
\begin{equation}
\label{eq:UphiPhim}
	U(\phi;\Phi_m)
    	%=\mathds{1}^{(1)}\otimes\mathds{1}^{(2)}\otimes\cdots\otimes U_1(\phi;\Phi_m) \otimes\cdots\otimes\mathds{1}^{(N)}
        = U_1(\phi;\Phi_m) \otimes\cdots\otimes\mathds{1}^{(N)}
\end{equation}
acting on the state space~(\ref{eq:rhoH2N}).

In a physical implementation of an interferometer, mechanical disturbances,
air-pressure changes
and the thermal fluctuations induce optical-path fluctuations. 
These effects randomize the phase difference
\begin{equation}
\label{eq:phasediffm}
	\phi-\Phi_m
\end{equation}
according to prior distribution~$p(\phi)$.
The quantum channel is thus~\cite{HS11b}
\begin{align}
	\mathcal{I}&(\phi;\Phi_m):
    	\mathcal{S}
        	\left(\mathscr{H}_2^{\otimes(N-m)}\right)
            \to\mathcal{S}
        		\left(\mathscr{H}_2^{\otimes(N-m)}\right):
            	\nonumber\\
		&\rho_m\mapsto
\int\limits_{\phi=0}^{2\pi}\text{d}\phi\,p(\phi) U^\dagger(\phi;\Phi_m)\rho_m U(\phi;\Phi_m),
\end{align}
where~$\rho_m$ is the state after the~$(m-1)^{\text{th}}$ photon is measured.

\paragraph{Detection, feedback, and inference.}
After the~$m^\text{th}$ photon passes through the interferometer, the photon is detected by one of the single-photon detectors positioned outside the output ports. 
The information of about the exit port is~$x_m\in\{0,1\}$,
which is given to a controller.
The controller then uses this information to compute~$\Phi_m$ from the policy~$\varrho$ before the next photon arrives.		
The procedure of simulating the injection of the next photon from the sine state~(\ref{eq:sinestate})
followed by action of the channel~(\ref{eq:UphiPhim})
and then measurement at the output ports
is repeated until all photons are consumed,
and the estimate is inferred from~$\tilde{\phi}=\Phi_N$,
assuming no loss of photons.

\paragraph{Imprecision}
Imprecision of the estimate~(\ref{eq:imprecision}) is related to the Holevo variance~\cite{WK97}
\begin{equation}
\label{eq:holevo}
	\left(\Delta\tilde{\phi}\right)^2
    	=V_\text{H}
        :=S^{-2}-1
\end{equation}
by the sharpness function
\begin{equation}
\label{eq:sharpness}
	S=\frac{1}{K}\left|\sum\limits_{k=1}^K
\exp\left[\text{i}(\phi_0^{(k)}-\tilde{\phi}^{(k)}\right]\right|,
\end{equation}
which quantifies the width of a distribution over a periodic variable.
The sharpness~(\ref{eq:sharpness}),
hence the Holevo variance~(\ref{eq:holevo}),
is estimated by repeatedly simulating~$K=10N^2$ times~\cite{HS10}
for uniformly randomly chosen~$\phi_0\in\left[0,2\pi\right)$ in each run
with~$\phi_0$ the unknown (noiseless) 
interferometric phase shift.
In the simulation~$\phi_0$ the mode
(most frequent value)
of the unimodal prior distribution $p(\phi)$.
The result $S$~(\ref{eq:sharpness})
is itself sampled from a distribution~$P_\text{sharp}(S)$
with a mean
\begin{equation}
\label{eq:averagesharpness}
	\bar{S}
    	=\sum SP_\text{sharp}(S).
\end{equation}
Consequently,
the Holevo variance~(\ref{eq:holevo}),
and thus the imprecision of the estimate,
is only estimated through this procedure.
\subsubsection{Policy generation}\label{subsubsec:policygen}
Whether the adaptive estimation scheme is capable of breaking the SQL and reaching HL depends on the feedback policy;
in general, the feedback policy does not come close to the HL although can surpass the SQL,
which is our success criterion.
We compare two set of policies,
one designed based on Bayes's theorem~\cite{BW00}
and the other by reinforcement learning~\cite{HS10,PWZ+17}.
The most significant difference between these two methods is that the Bayesian feedback computes~$\Phi_m$ based on a trusted model of how the quantum state evolves,
whereas the reinforcement-learning algorithm here uses only the measurement outcomes~$x_m$ to decide the value of~$\Phi_m$.
A model-free approach that is used in the reinforcement-learning algorithm should, in principle, be robust to changes in the estimation scheme although whether the policies can preform better than the model-based approach is unknown.
\paragraph{Bayesian feedback.}
\label{para:bayesianfeedback}
The idea behind the Bayesian feedback is that every outcome from the measurement improves our knowledge of the the value of~$\phi_0$~\cite{BW00,BWB01}. 
This is mathematically captured in Bayes's theorem~\cite{RHB06_ch30},
where the initial knowledge of~$\phi_0$ is represented by a uniform prior, indicating that~$\phi_0$ can be any value in~$\left[0,2\pi\right)$ with equally probability. 

For each particle exiting the interferometer, the probability for the particle being detected in output port~$x_m$ can be computed 
by assuming a perfect input state and known quantum dynamics such as the unitary evolution~(\ref{eq:UphiPhim}).
The prior for~$\phi_0$ is then updated using Bayes's theorem. 
The squared width of this prior is quantified by~$V_\text{H}$ (\ref{eq:holevo}) in each measurement step;
the optimal~$\Phi_m$ that minimizes width for the next particle is then computed from the model.

Although the Bayesian method gets close to the HL~\cite{BWB01},
the problem with Bayesian inference subject to a trusted model is that the policy might not be robust to variations of input state and to noise in the system.
This mismatch is known to be a major concern in designing controllers~\cite{Lei04_ch6,WSB+16} and one way to address this concern is to adopt a design-and-feedback approach that does not use models but the data from the physical setup~\cite{HW13}.
\paragraph{Reinforcement learning.}
\label{para:reinforcementlearning}
Reinforcement learning is used to devise a policy~$\varrho$ by trial and error~\cite{SB2017_ch1,SBW92},
and the learning algorithm iteratively
optimizes~$\varrho$ by testing the policy based on evaluating the outcome of the specified control task.
For noisy adaptive phase estimation,
policy evaluation is based on average sharpness (\ref{eq:averagesharpness})~\cite{HS11b},
which averages out noise.
In practice,
due to the high cost of sampling $P_\text{sharp}(S)$~(\ref{eq:averagesharpness})
very few runs are performed to obtain~$\bar S$.

Reinforcement learning does not employ a model but rather uses the control outcomes to learn~\cite{DPS12};
however,
we currently train based on simulations that employ a model,
but training could and should ultimately be performed under conditions of deployment such as in the laboratory or in the field.
For reinforcement learning,
feedback is assumed to follow a logarithmic-search heuristic Markovian update rule~\cite{HS10}
\begin{equation}
\label{eq:rl update}
	\Phi_{m-1}\mapsto\Phi_{m-1}-(-1)^{x_m}\Delta_m,
\end{equation}
\label{eq:Delta}
with the phase-adjustment vector
\begin{equation}
\label{eq:phaseadjustmentvector}
	\bm{\Delta}
    	:=\left(\Delta_1,\Delta_2,\dots,\Delta_N\right),
\end{equation}
optimized during the training stage by a scalable, noise-resistant reinforcement-learning algorithm such that average sharpness~(\ref{eq:averagesharpness}) is maximized~\cite{PWZ+17}.
This update rule~(\ref{eq:rl update})
corresponds to turning the ``phase knob''
up or down by a fixed amount~$\Delta_m$,
after the $m^\text{th}$ photon,
subject only to the previous outcome and ignoring the full measurement history.

Reinforcement learning incurs a time cost to generate a policy~$\varrho$. 
The time cost for generating the policy is quantified by loop analysis of the learning algorithm, assuming the algorithm is run on a single processor~\cite{LCPS13}.
The scaling of time cost with respect to the particle number $N$ determines the complexity, and in practice this scaling is polynomial so the degree of the polynomial convey the complexity for generating the policy.

\subsection{Models of phase noise}
\label{subsec:noisemodels}
In this subsection, we explain the choices of the phase-noise model for the robustness test.
This noise is simulated turning~$\phi$ into a random variable that has a unimodal probability distribution with the peak at~$\phi_0$. 
The mode~$\phi_0$ is assumed to be the unknown parameter to be estimated. 
For the test,~$\phi$ follows one of these four distributions:
normal,
three-stage random telegraph,
skew-normal or
log-normal distribution.
We summarize the relationship between the noise parameters and the variance and skewness~\cite{Sev05_ch4} as we use both in selecting the parameters for the robustness test and the variance in particular to quantify the noise level.
\subsubsection{Normal-distribution noise} 
Normal-distribution noise is important for testing robustness of the learning algorithm because the normal distribution is especially prominent due to the central-limit theorem, which states that the average of a random variable has a normal distribution~\cite{Sev05_ch12}. 
Due to the prevalence of the normal distribution,
assuming normal-distribution noise model is common~\cite{Kir12_ch3}.
The normal distribution
\begin{equation}
\label{eq:normal-distribution noise}
	p(\phi)
    	=\frac{\text{e}^{\frac{(\phi-\mu)^2}{\sigma^2}}}{\sqrt{2\pi}\sigma},
\end{equation}
is parametrized by the
mean~$\mu$ and standard deviation~$\sigma$. 
As this distribution is symmetric,
skewness~$\gamma$ is identically zero and thus the mode is at~$\mu$,
and the variance is~$V=\sigma^2$. 

In our simulations,
we set~$\mu\equiv\phi_0$
so the only free parameter is~$\sigma$, which is bounded above by~$\sigma<\pi$ as otherwise the width would exceed the domain of~$\phi$. This value of~$\sigma$ is, of course, too high to expect any measurement scheme to operate with reasonable imprecision, and so the robustness test is conducted lower~$\sigma$.

\subsubsection{Random telegraph noise}
Random telegraph noise~\cite{New68} is a discrete distribution that, for each time step,
randomly switches between two values, one being the correct and the other an erroneous value.	
Whereas this noise is most relevant to digital electronics as it simulate a bit-flip error, 
it can be use to simulate other digitized noise, such as salt-and-pepper noise in image processing~\cite{Bon05}. 

We modify two-stage random telegraph noise to have three stages,
\begin{align}
	p(\phi)=
\begin{cases}
1-p_{\text{s}}, & \phi=\phi_0,\\
\frac{p_{\text{s}}}{2}, & \phi=\phi_0\pm\delta.
\end{cases}
\end{align}	
The probability of switching to an erroneous value is~$p_{\text{s}}$, and~$\delta$ is the distance between the true and erroneous values leading to
\begin{equation}
\label{eq:rtnoiseVgamma}
	V=p_{\text{s}}\delta^2,\;
    \gamma\equiv0
\end{equation}
with the last relation following from the symmetry of the distribution.

Unimodality of the distribution implies that~$p_{\text{s}}<2/3$.
Furthemore, $\delta<\pi$ to avoid overlap with another distribution over the phase domain.
To comply with both constraints and being able to raise the noise level to at least~$V=3$ so the result can be compared to that of other distributions,
we fix~$p_\text{s}$ for the test and vary only~$\delta$. 

\subsubsection{Skew-normal-distribution noise}
The skew-normal distribution~\cite{AC14} is modified from a normal distribution by multiplying with a function whose skewness parameter is~$\alpha$.
Skew-normal noise is a class of noise that includes normal-distribution noise as a limiting case.
Although this distribution is not widely used as noise, it arises in simulations of noise for filters and detectors~\cite{Pou07,RE14}.

The	skew-normal distribution is
\begin{equation}
\label{eq:snd}
	p(\phi)
    	=\frac{\text{e}^{-\frac{(\phi-\mu)^2}{2\sigma^2}}}{\sqrt{2\pi}\sigma}\left[1+\text{erf}\left(\frac{\alpha}{\sqrt{2}\sigma}(\phi-\mu)\right)\right]
\end{equation}
for~$\text{erf}()$ the error function~\cite{RHB06_ch18}.
Skewness of the distribution is
\begin{equation}
\label{eq:skewnessskewnormal}
	\gamma=\frac{4-\pi}{2}
    	\frac{2\beta}{\pi-2\beta},\;
	\beta=\frac{\alpha^2}{1+\alpha^2},
\end{equation}
and the variance is 
\begin{equation}
V=\sigma^2\left(1-\frac{2\beta}{\pi}\right)
%	V=\sigma^2\left(1-\frac{2\alpha^2}{\pi(1+\alpha^2)}\right).
\end{equation}
The mode, however, does not have a closed form although it remains close to $\mu$ as $\alpha/\sigma$ increases. For the simulation, we assume the mode is~$\mu$.

\subsubsection{Log-normal-distribution noise}
Log-normal~\cite{LSA01} noise has a heavy-tailed skewed distribution that provides another approach to generalizing the normal distribution and is employed in the study of networks~\cite{KGDK2015,KSF87}
and electronics~\cite{Jou11}.
In this case, the logarithm of the random variable is said to have a normal distribution, leading to the distribution 
\begin{equation}
	p(\phi)
    	=\frac{\text{e}^{-\frac{(\log \phi -\mu')^2}{2\sigma'^2}}}{\sqrt{2\pi}\sigma'\phi}
\end{equation} 
with mode and variance
\begin{equation}
\label{eq:phi0lognormal}
	\phi_0=\text{e}^{\mu'-\sigma'^2},\;
    V=\left(\text{e}^{\sigma'^2}-1\right)\text{e}^{2\mu'+\sigma'^2},
\end{equation}
respectively, and skewness
\begin{equation*}   
	\gamma
    	=\left(\text{e}^{\sigma'^2}+2\right)\sqrt{\text{e}^{\sigma'^2}-1}.
\end{equation*}
As this distribution is defined for~$\phi\in\left(0,\infty\right)$, we first generate a random number within the compact phase domain given~$\mu'$ and~$\sigma'$ 
and then apply the shift
\begin{equation}
	\phi\mapsto\phi
    	+\phi_0-\text{e}^{\mu'-\sigma'^2}
\end{equation}
so that the mode of the distribution is centred at $\phi_0$~(\ref{eq:phi0lognormal}).

\subsection{Regression analysis}
\label{subsec:regression}
The imprecision~$\Delta\tilde{\phi}$ and~$N$ are asymptotically power-law related~(\ref{eq:imprecision}).
However, when the system is noisy,
this relationship fails for low~$N$,
with the actual bound on~$N$ depending on the noise model.
We employ regression analysis to select the subset of~$V_\text{H}$ at high~$N$ that scales as~$N^{-\wp}$
and estimate the corresponding~$\wp$ 
by building piecewise functions and selecting the best candidate to represent the data.
In this subsection,
we explain our regression-analysis procedure
for fitting a model given a set of data.
\subsubsection{Fitting the model}
Regression analysis aims to determine the mathematical relationship between dependent
($V_\text{H}$ here)
and independent variables
(here~$N$)~\cite{MPV12_ch1}.
The process of building this mathematical model begins with selecting a function~$f(N)$
based on the knowledge of the mechanism and observations of the trends~\cite{CS2013_ch2}.
The function is only a best guess as the discerned trend could be subjective.

After a function is selected, the function is then fitted to the data
by finding the parameters that minimize the error between 
the predicted $V_\text{H}$ and the data $V_\text{H}$~\cite{CH12_ch1}.
The method we employ is the least-squaress estimation~\cite{MPV12_ch2},
used in linear regression to calculate the variables 
by constraining the gradients to zero and solving the resulting system of linear equations.
We choose this method as we fit linear and piecewise linear equations to log-log plot of $V_\text{H}$ and $N$.
\subsubsection{Consistency of the fit}
\label{subsubsec:criteria}
As the fitted function is only an educated guess,
the fitting result must be examined for inconsistencies with respect to the model's assumptions~\cite{CH12_ch1}.
An alternative function can then be proposed, fitted, and compared to the previous function
in order to find one that best represents the data.
Deciding on the best model from the set is done using statistical criteria 
that either estimate the goodness of the fit to the data or between two models fitted to the same data~\cite{MPV12_ch10}.
The model that is consistently shown to fit well according to each of the criteria
is then selected to represent the data.

Common linear-regression criteria include
\begin{itemize}
\item	the coefficient of determination
	\begin{equation}
		R^2
        	=1-\frac{\sum\limits_N 	
            	\left(V_\text{H}^{(N)}-
                	f(N)\right)^2}{\sum\limits_N 
                    	\left(V_\text{H}^{(N)}-\bar{V}_\text{H}\right)},
\end{equation}
	adjusted to
    \begin{equation}
    \label{eq:adjusted R-squared}
		\overline{R^2}
        	=R^2-\frac{b}{v-b-1}(1-R^2),
	\end{equation}
with~$v$ the number of data points in the fit.
\item	the corrected Akaike Information Criteria $\text{AIC}_\text{c}$,
		which quantities information lost due to the discrepancy between the model function and the true function $g(N)$, and is `corrected' to avoid overfitting,
\item	the~$F$-test, which assesses a full model (maximum~$b$),
		as the null hypothesis,
        vs a reduced model (reduction from the full model) as the alternative hypothesis~\cite{CS2013_ch1,CS2013_ch2}, and
        
\item	Mallows's $C_p$~\cite{CS2013_ch2,MPV12_ch10} 
		(but we use~$b$ rather than the traditional~$c$ for the number of parameters),
        which estimates the mean-square prediction error~\cite{MPV12_ch10}
to compare a reduced model to the full model, where the reduced model with the smallest~$C_p$ close to $b$ is chosen.
\end{itemize}
Each of these criteria is designed to penalize functions 
with many parameters~$b$ to avoid overfitting the data~\cite{CS2013_ch2}.

\section{Approach}
\label{sec:approach}
In this section,
we devise a test to determine whether quantum-enhanced precision is feasible in the presence of unknown noise.
We then assess whether power-law scaling of phase imprecision vs particle number~$N$
is valid asymptotically and establish a method to determine this power~$\wp$.
Finally,
we define the resource for generating and implementing the control policies in terms of the scaling of the space and time cost with $N$.

\subsection{Robustness test}
\label{subsec:robustnesstest}
The robustness of AQEM policies is determined by testing the policies in the presence of noise whose model is not recognized by the policies and the method that generates the policies,
although reinforcement-learning policies are learned in trainings that include the noise.
Here we define the test for adaptive phase estimation, including phase noise from~\S\ref{subsec:noisemodels}.
We specify the domain of~$N$
for simulating the phase estimation schemes to obtain~$V_\text{H}$ in noisy conditions.
The noise parameters are variance~$V$ and skewness~$\gamma$ (\S\ref{subsec:noisemodels}),
but here we fix~$\gamma$ for the asymptotic distributions,
and we obtain the robustness-test threshold
in terms of~$V$, which is the maximum for each noise model such that the SQL is violated.

\paragraph{Varying~$N$.}
To ascertain the asymptotic value for~$\wp$,
we simulate adaptive phase estimation for
\begin{equation}
	N\in\{4,5,\dots,100\},
\end{equation}
as~$V_\text{H}$
computed from this domain is sufficient to show power-law relationship at high~$N$.
Furthermore, increasing~$N$ further 
requires changing double-precision arithmetic to quadruple-precision arithmetic to generate and manipulate the sine state without rounding error.
Consequently, this increase in precision leads to a fifteen-fold increase in run-time at $N=100$, 
which is a large expense for generating a single data point. 
Therefore, we do not attempt to verify the robustness beyond this 100 particles.
\paragraph{Skewness.}
\label{para:skewness}
We fix skewness~$\gamma$ to a single value for all runs and only vary~$V$
because~$V$ is the dominant term in our noise models and~$\gamma$ has a small effect~\cite{PWS17}.
We fix the skewness for the asymmetric distribution to
\begin{equation}
\label{eq:gammma=0.8509}
	\gamma=0.8509,
\end{equation}
which is sufficiently large to distinguish between the various noise models;
otherwise all noise looks Gaussian.
This value of $\gamma$~(\ref{eq:gammma=0.8509})
corresponds to $\alpha=5$ for the skew-normal distribution 
where we are able to observe its effect on~$\wp$ when compared to symmetric noise distributions. 
This same level of skewness corresponds to $\sigma'=0.2715$ in the log-normal distribution.

\paragraph{Robustness threshold.}
Our policy is robust if the SQL-breaking
condition~$\wp>1/2$ is satisfied for all four noise models in~\S\ref{subsec:noisemodels}.
As discussed in~\P\ref{para:skewness},
we fix~$\gamma$,
and we ignore higher cumulants;
thus the policy robustness threshold is in terms of~$V$,
i.e., the maximum~$V$ such that $\wp>1/2$ holds
for all four noise models.
This optimization problem is hard
so we adopt a simpler characterization procedure instead to get insight into the robustness threshold.
Our approach is to run the simulations 
for~$V\in\{1,2,3\}$ for symmetric noise and~$V\in\{1,3,5,7\}$ for asymmetric noise,
and we do not push beyond $V=7$ to keep below an imprecision width of $2\pi$.
We use these data to determine whether AQEM policies pass the robustness test.

\subsection{Determining asymptotic power-law scaling}
\label{subsec:regressionprocedure}
To ascertain the robustness of AQEM policies, 
the asymptotic~$\wp$ is estimated from a subset of $V_\text{H}$ at sufficiently high $N$,
and determining this subset is done by fitting piecewise linear equations to a log-log plot of $V_\text{H}$ vs $N$.
In this subsection, we introduce five piecewise functions that are constructed from observations regarding the trend of  $V_\text{H}$ vs $N$.
We then explain the method of finding the break points between segments in the piecewise function and fitting the functions to the data.
Using the criteria~\S\ref{subsec:regression},
we create a majority-vote method for selecting the function that best represents the data and thus~$\wp$ from the last segment of the fit is used to estimate the asymptotic scaling.

\subsubsection{Piecewise models.\label{subsub:piecewise models}}
The trend in $\log V_\text{H}$ vs~$\log N$ differs under noisy conditions, 
and here we describe the trends we have observed that lead to piecewise~linear functions.
We construct five such functions, containing 1 to 3 segments that are then fitted to $V_\text{H}$ vs $N$.

When the interferometer is noiseless, the relationship appears to be a power-law captured in a linear equation,
although the accept-reject criterion in the reinforcement-learning algorithm can lead to a different~$\wp$ for $N>93$.
Once the noise level becomes high, typically $V>1$, the relationship does not appear to be linear for low $N$.
Therefore, we include segmented models that fits linear interpolation to the data in the first segment.

Combining these observations, we construct five piecewise-linear models that can potentially represent $\log V_\text{H}$ as a piecewise function of $\log N$.
The models have 1 to 3 segments,
each segment connected at the break points 
determined by the fitting methods.
Three of the models are one, two, and three linear models,
whereas the other two are a two-segment model, 
where the first segment (low~$N$) is a linear interpolation, 
and a three-segment model where the first segment is a linear interpolation and the second and third segments are linear.

\subsubsection{Fitting method}
\label{subsubsec:fittingmethod}
The method for fitting the linear equations that we use is the least-squares method~\cite{MPV12_ch2}.
However, because the functions in~\S\ref{subsub:piecewise models} are segmented, 
we include a step to optimize the break points depending on the specific function.

The full model for the regression analysis is the three-segment linear function, 
which is fitted using linear-square method and the segments 
determined by a heuristic global optimization algorithm~\cite{Jek17}.
The two-segment linear function is also fitted using the same least-square method 
although a brute-force search is used to find the break point starting from $N=4$.

The method for finding the break points for models with interpolation are different
 as the linear interpolation leads to a small residual.
 Thus, optimizing 
using the least-squares method 
can lead to a single segment of linear interpolation.
For this reason, we first find the stop point for the first segment 
by fixing the latter segments to a single linear line 
and search for break point that results in a large decrease in sum square error.
As for the single-segment linear model, we use a standard library to fit to the data.

\subsubsection{Fitting figures of merit and model selection}
\label{subsubsec:modelselection}
After the functions are fitted to $V_\text{H}(N)$, the criteria
$\overline{R^2}$, $\text{AIC}_\text{c}$, 
the $F$-value, and Mallows's $C_p$ (\S\ref{subsubsec:criteria}) are calculated for each of the function
and the fits are visually inspected.
These criteria are used to select the function that best fits the data.
Here we explain how the best function is chosen.

After the functions are fitted and the criteria are calculated,
each fit is visually presented and inspected to ascertain that the the segmentation fits the pattern.
If correction are unnecessary, 
the functions are then ranked for each of the criteria.
Note that we do not perform the full $F$-test as we discovered that reduced models typically fail the test even though there is no discernible difference when compared to the full model.
However, the $F$-value can still be used to quantify the difference between using the full and reduced model,
so we use the $F$-value to rank the functions instead of conducting a pass-fail test.

After the functions are ranked, the function that is voted as best by most of the criteria is chosen to represent the data.
In the case where the full model, the three-segment linear function, is voted according to the criteria,
the value of~$\wp$ from the last segment and the subset of $N$ where this value is computed 
is compared to the next alternative function to determine whether the function overfits the data.
If~$\wp$ from the two functions differs more than 0.001, then the full model is chosen;
otherwise, the alternative function is chosen.
The limit we use here is specified based on the precision used in this paper
and can be changed based on the desired precision of~$\wp$.

\subsection{Resource complexity}\label{subsec:resourcescaling}
% * <panpalitta@gmail.com> 2018-09-13T00:29:36.719Z:
% 
% This subsection needs checking
% 
% ^.
To compare and select between policies and methods of generating policies, we determine the complexity of designing and implementing a policy using the loop-analysis method in algorithm analysis.
We begin this subsection by explaining the time complexities of generating policies.
We then explain what the controller does and the requisite resources, quantified by the space and time complexity for executing a policy with~$N$ particles.

\subsubsection{Design complexity}
When an optimization or a learning algorithm is used to generate a policy, there is a time cost associated with the use of the algorithm.
The scaling of the upper bound of this cost is called the design complexity. 
Here we explain the assumptions behind the calculation of this complexity.

We assume that this task is performed on a simulation of the AQEM task, as is common practice in policy generation in quantum control~\cite{KRK+05},
and, therefore, the time cost includes the cost of simulating the AQEM task.
We assume that only a single processor is used for the purpose of comparing policies generation methods, although this cost can be reduced by parallelizing the learning or the optimization on multiple CPUs.

The design complexity for the reinforcement-learning policies in~\S\ref{subsubsec:policygen} is shown to be $O(N^6)$ through loop analysis~\cite{LCPS13},
and this complexity does not change when noise is included.
For policies that are devised through analytical optimization, 
such as the Bayesian feedback, 
this cost is zero as no algorithm is used. 

\subsubsection{Controller complexities}
\label{subsubsec:policycomplexity}
In this subsection, we explain the resource complexity required to use AQEM policies, quantified by the scaling of the space and time costs with the number of particles $N$~\cite{VK14_ch1}.
We begin by explaining the connection between an AQEM policy and an algorithm by viewing the controller as a computer, allowing us to use the method of algorithm analysis to calculate the complexities~\cite{VK14_ch5}. 
We then define the space and time cost for implementing policies and how these costs are calculated.

\paragraph{Controller}
The controller holds a policy
$\varrho$~(\ref{eq:policyupdate})
and uses this policy to execute decisions based on feedback from measurement of where the particles are detected.
To execute this task 
the controller requires computer memory and
sufficient time to affect the policy, 
which we use to determine the scaling of resource cost.

The controller is essentially a computer that receives input from detectors and transmits a control signal to an actuator that shifts the interferometric phase.
In this perspective,
the policy~$\varrho$
is represented as a computer algorithm expressed as a computer program. 
The candidate policy can be generated by various means including Bayesian feedback~\P\ref{para:bayesianfeedback}
and reinforcement learning~\P\ref{para:reinforcementlearning}.
Space and time costs are discussed in the next two subsubsections.

\paragraph{Space complexity}
We determine the upper bound for space cost,
which is the worst-case amount of memory used by an algorithm reported as a big O function of the size of the problem~\cite{VK14_ch1}.
As~$\varrho$ is executed by an algorithm,
this worst-case,
or maximum-size,
memory corresponds to how much space is required to hold the critical information required
to execute the feedback.

For~$\varrho$,
the computer's space cost for memory
depends on the type of policy.
For Bayesian feedback,
the size of the stored policy is~$O(N^2)$,
as shown in~\S\ref{subsec:bounds} and specifically in Table~\ref{tab:complexity},
and the size of the stored policy for reinforcement learning is $O(N)$~\cite{LCPS13}.
This linear scaling for policies obtained by reinforcement learning
is due to the generalized-logarithmic-search heuristic, leading to the size~$N$ 
phase-adjustment vector~(\ref{eq:phaseadjustmentvector}).
The information used by the policy are the policies parameters, where there are $N$ for an AQEM scheme that uses $N$ particles.

\paragraph{Time complexity}
\label{subsubsec:time}
Time complexity is the scaling for the upper bound in time cost for implementing a single shot of AQEM. 
This cost is calculated by assuming the time a particle takes to pass through an interferometer is constant,
mimicking the physical implementation of the control procedure, and use loop analysis, which counts the number of loops that perform operations,
which we assume all take the same constant time~\cite{VK14_ch5}.
For a shot of AQEM task using $N$ particles, $\Phi_m$ is computed $N$ times, corresponding to each particle passing through the interferometer and being detected.
For each particle, there can be loops nested in the computation of $\Phi_{m+1}$ according to the policy $\varrho$.
The complexity is reported as the scaling of this time cost with $N$.

We determine the implementation cost for reinforcement-learning policies by recognizing that the update of $\Phi_m$ according to Eq.~(\ref{eq:rl update}) is constant in time.
Therefore, the adaptive phase estimation procedure consisted of only one loop over the number of particles, and the implementation complexity is $O(N)$.
Bayesian feedback, on the other hand, has nested loops for updating the quantum state, which is $O(N^2)$ in time complexity for every computation of $\Phi_m$.
Hence, the implementation complexity is $O(N^3)$.

\section{Results}
\label{sec:results}
In this section, we report results for the robustness test and the resources
based on sampling from simulated adaptive phase estimation.
We present and compare
$V_\text{H}(N)$~(\ref{eq:holevo})
obtained by reinforcement learning discussed in~\P\ref{para:reinforcementlearning}
and by Bayesian feedback discussed in~\P\ref{para:bayesianfeedback}.
Our analysis considers all four types of noise discussed in~\S\ref{subsec:noisemodels}.
We report
values of $\wp$~(\ref{eq:imprecision})
from the regression procedure discussed in~\S\ref{subsec:regressionprocedure}
and resource complexities discussed in~\S\ref{subsec:resourcescaling} for both the reinforcement-learning policies and the Bayesian feedback.

\subsection{Variance vs number of particles}
\label{subsec:variancen}
In this subsection, we present results for~$V_\text{H}$ as a function of number~$N$ of particles.
Specifically, 
we present plots of~$V_\text{H}$ vs~$N$
from~4 to~100 particles,
which is enough to determine scaling as discussed in~\S\ref{subsec:resourcescaling}.
Both cases of using reinforcement-learning policies and Bayesian feedback are presented as log-log plots and compared to the SQL,
with these plots obtained by computing
from simulations of noiseless phase estimation using a product state $\ket{0,1}^{\otimes N}$ following the notation of Eq.~(\ref{eq:symmetricbasis}). 
The HL is generated based on the intercept of the SQL data using the scaling of $1/N^2$ to provide a benchmark. 

\subsubsection{Reinforcement learning}
\label{subsubsec:RLvariance}
\begin{figure*}
	\subfloat[normal-distribution noise]{\includegraphics[width=0.45\textwidth]{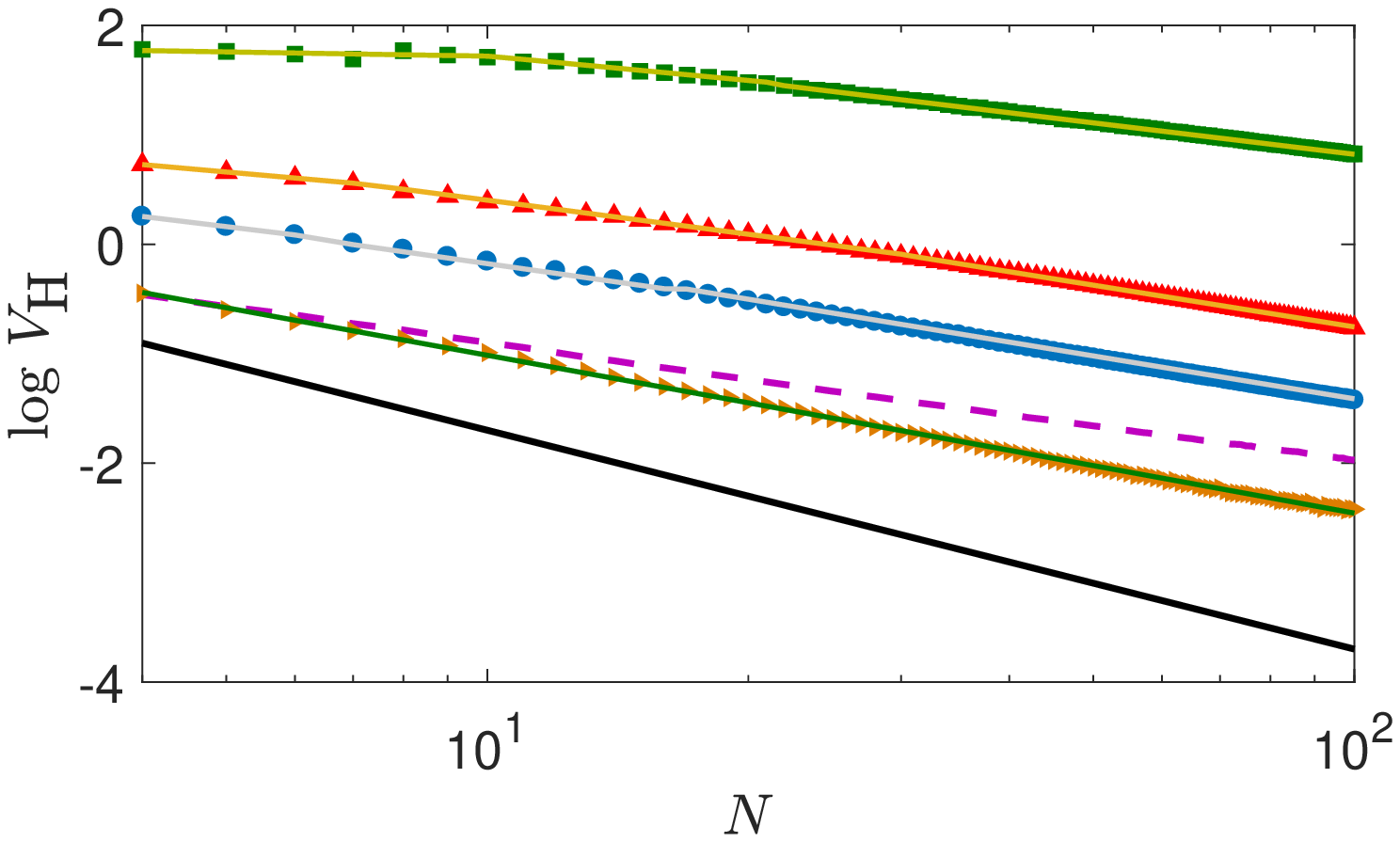}}
	\subfloat[random telegraph noise]{\includegraphics[width=0.45\textwidth]{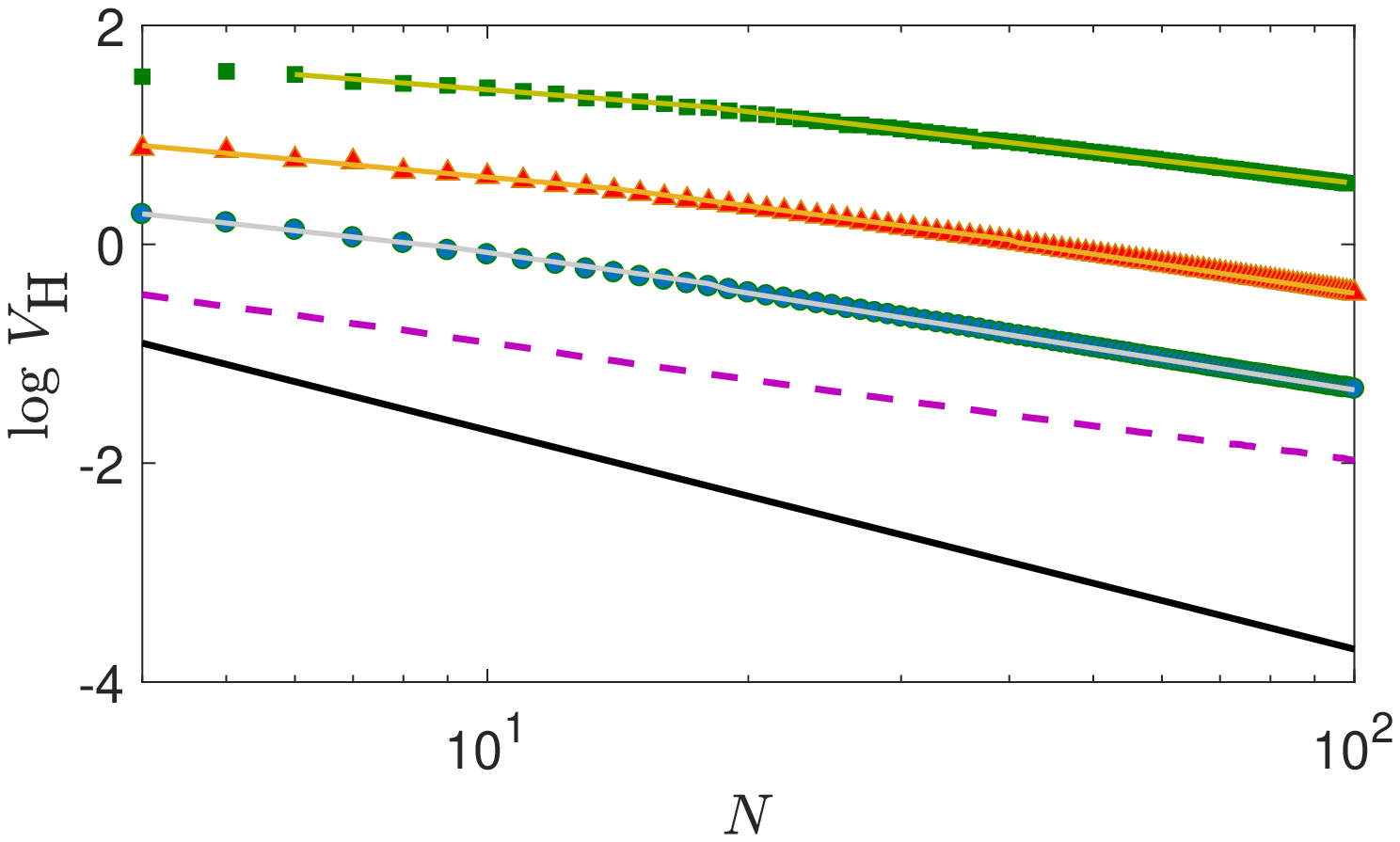}}
	\\
	\subfloat[skew-normal-distribution noise]{\includegraphics[width=0.45\textwidth]{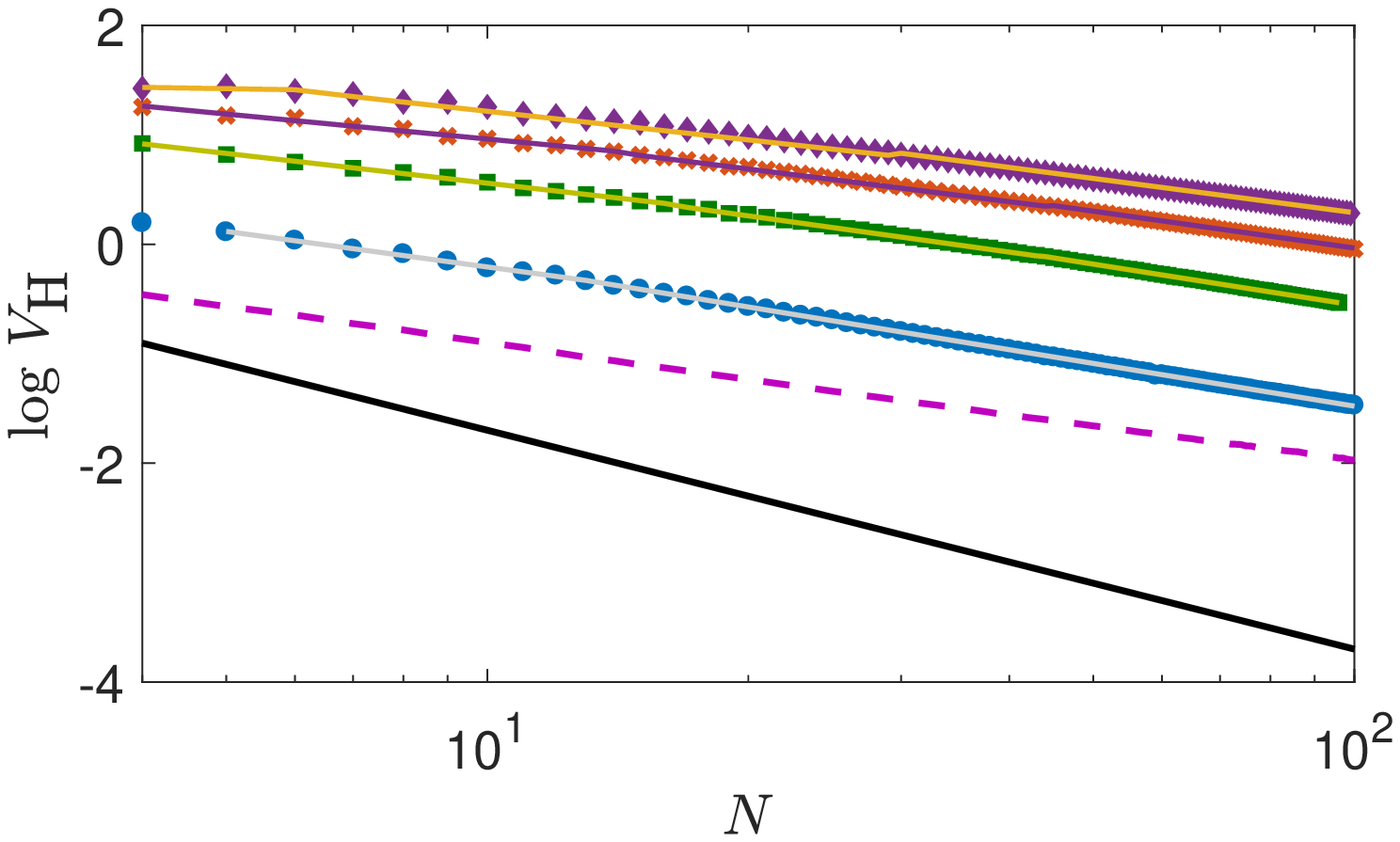}}
	\subfloat[log-normal-distribution noise]{\includegraphics[width=0.45\textwidth]{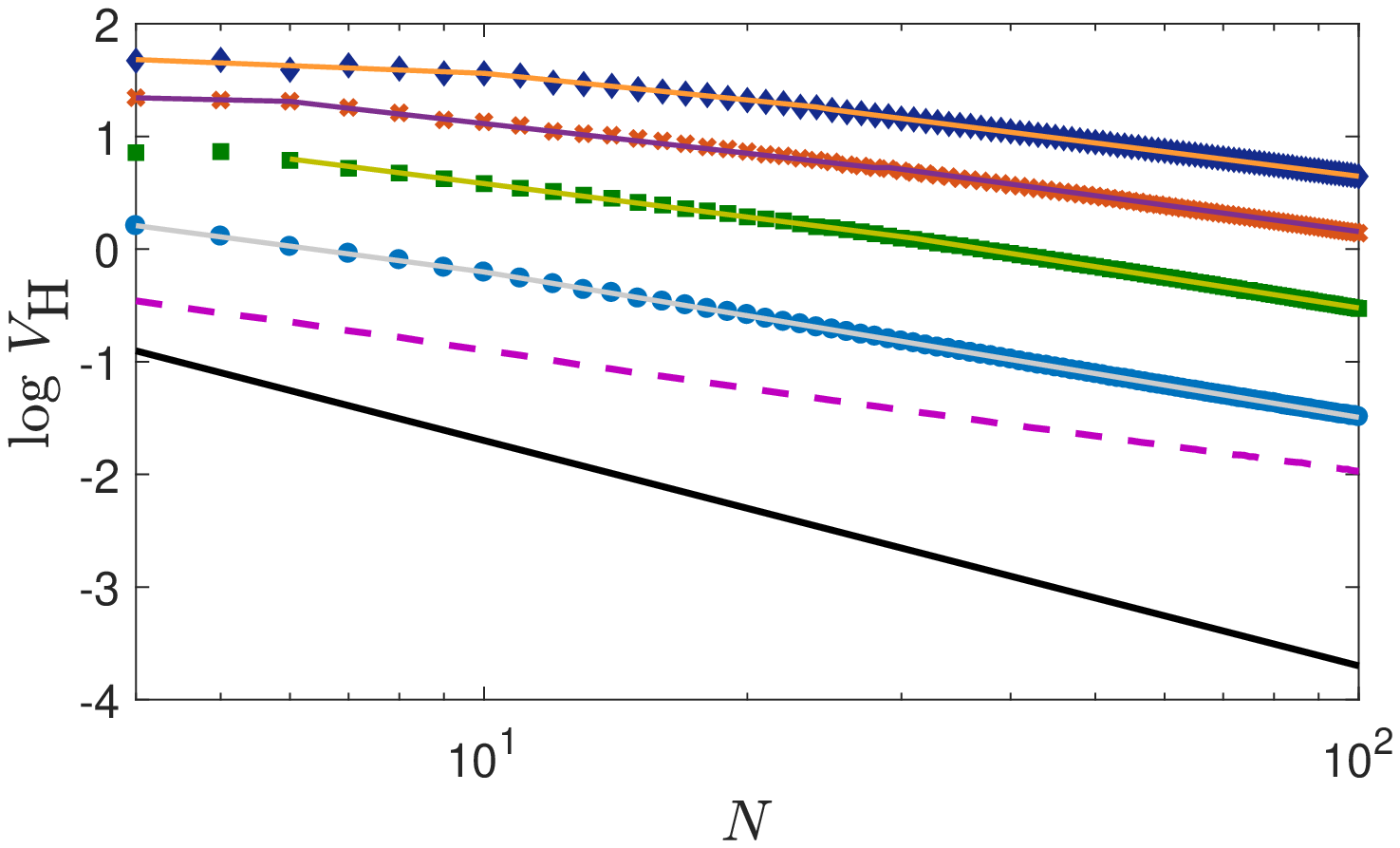}}
	\caption{\label{fig:noise data RL}%
		Logarithmic plots of Holevo variance from simulation of adaptive phase estimation. The policies are generated using reinforcement learning implemented in the specified noise condition, namely,
        (a) normal-distribution noise,
        (b) random telegraph noise,
        (c), skew-normal-distribution noise, and (d) log-normal-distribution noise.
		The plot for the normal-distribution noise also includes the data from the noiseless simulation (brown side-facing triangle) and its linear fit (green solid).
		The blue circles are data when $V=1$, the red triangles when $V=2$, the green squares when $V=3$, the brown plus when $V=4$, the brown crosses when $V=5$ , and the purple diamonds when $V=7$. 
The lines shown are the piecewise~linear fits of the data whose scaling is reported.
		The solid black line if the HL and the dashed purple line in the SQL generated from noiseless adaptive phase estimation.}	
\end{figure*}

Here we present the log-log plots of $V_\text{H}$ vs~$N$ from adaptive phase estimations using reinforcement-learning policies,
as shown in Fig.~\ref{fig:noise data RL}.
Subfigures~\ref{fig:noise data RL}(a--d)
present~$V_\text{H}$ when normal-distribution noise, random telegraph noise, skew-normal noise, log-normal noise are included,
respectively.

Figure~\ref{fig:noise data RL}(a) also includes~$V_\text{H}$ from noiseless interferometry. This locus appears as a straight line in the plot, indicating a power-law relationship between~$V_\text{H}$ and~$N$.
As the noise variance~$V$ increases,
this power-law relationship breaks into two parts, clearly visible in the~$V_\text{H}$ vs~$N$ plot
from~$V=3$.
This trend also appears at~$V=2$ as the
the model selection procedure~\ref{subsubsec:modelselection} selects the two-segment model for this data set.
The observation that the power-law relationship fails when noise is included is also evident in Figs.~\ref{fig:noise data RL}(b--d).
In these cases, the plots are fit to two- or three-segment linear equations as~$V_\text{H}$ appears to have a bump at low~$N$ as~$V$ increases.

The increase in phase noise~$V$ also results in an increase in the intercepts of~$V_\text{H}$ power-law lines;
however, the rate of change appears to depend on the noise model.
The difference can be seen in Fig.~\ref{fig:noise data RL}(a--b),
both include symmetric noise distributions but with different spacing of the intercepts.
The same observation holds for Fig.~\ref{fig:noise data RL}(c--d), which are from asymmetric distributions.
Comparing the four plots shows that the intercepts appear to increase slower for asymmetric distributions than the symmetric distributions, being close to 1 for $V=3$ in the former and $V=2$ 
for the latter.

\subsubsection{Bayesian feedback}
\label{subsubsec:BFvariance}
\begin{figure*}
	\centering
	\subfloat[normal-distribution noise]{\includegraphics[width=0.45\textwidth]{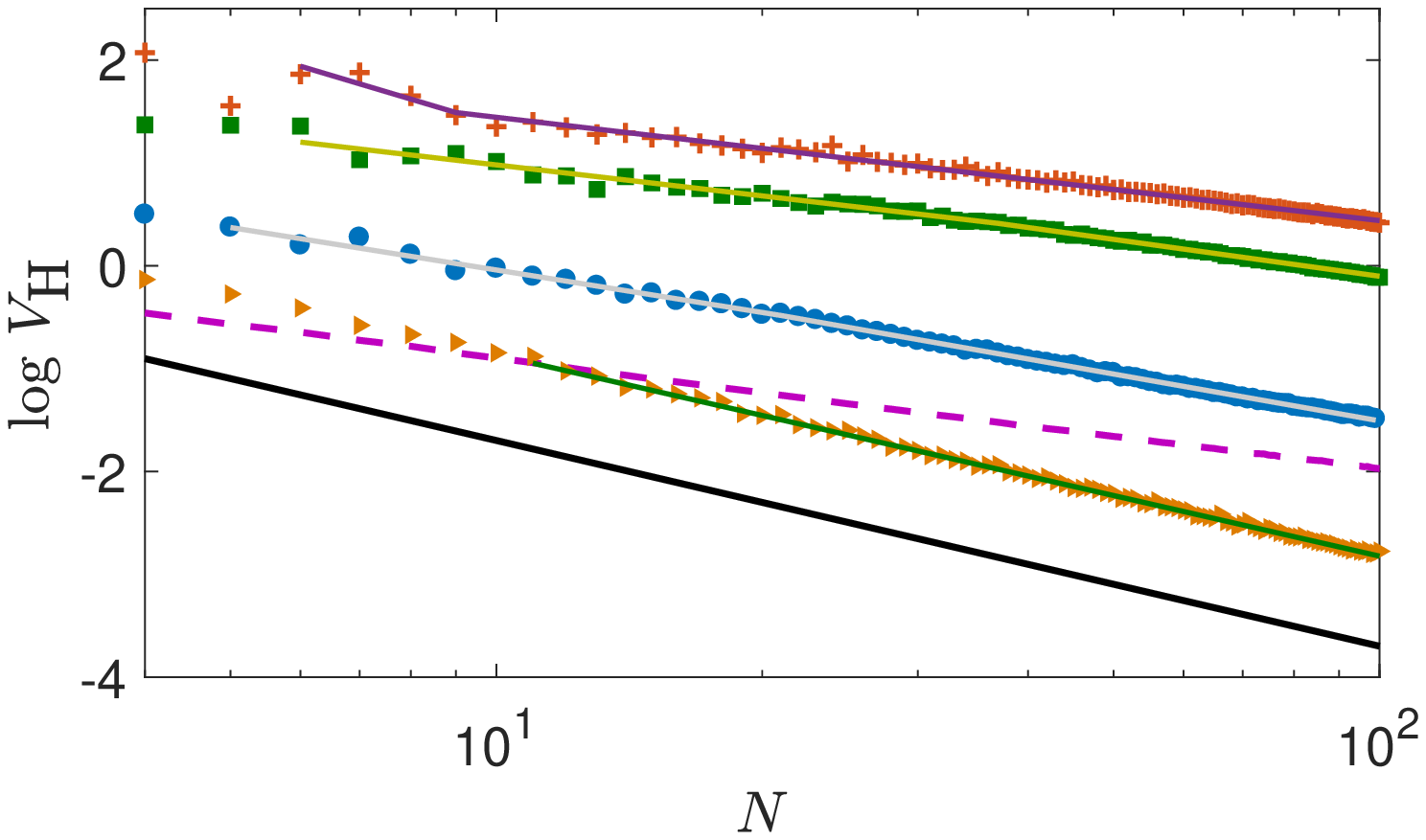}}
	%(a) normal-distribution noise\\
	\subfloat[random telegraph noise]{\includegraphics[width=0.45\textwidth]{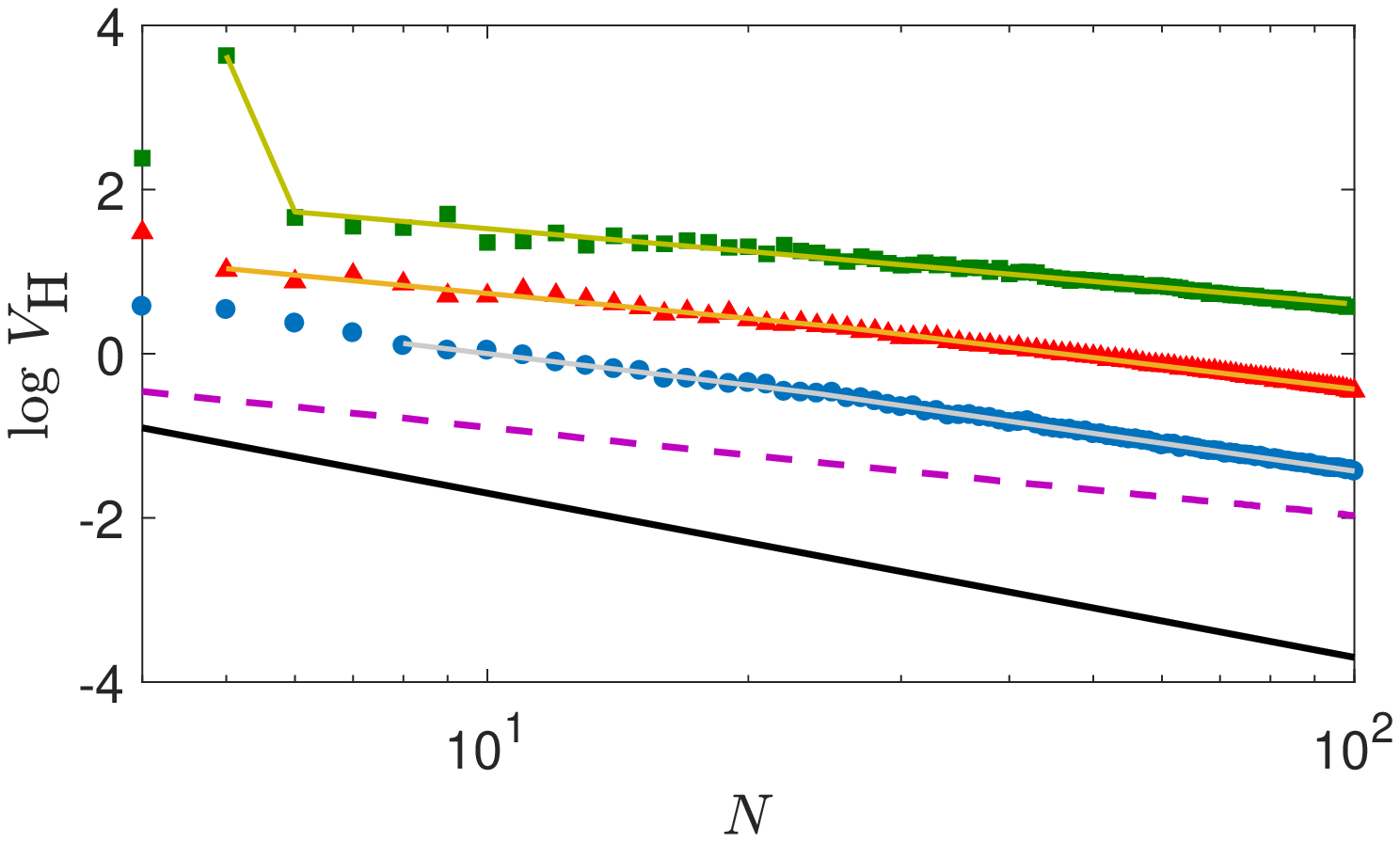}}
	%(b)random telegraph noise
	\\
	\subfloat[skew-normal-distribution noise]{\includegraphics[width=0.45\textwidth]{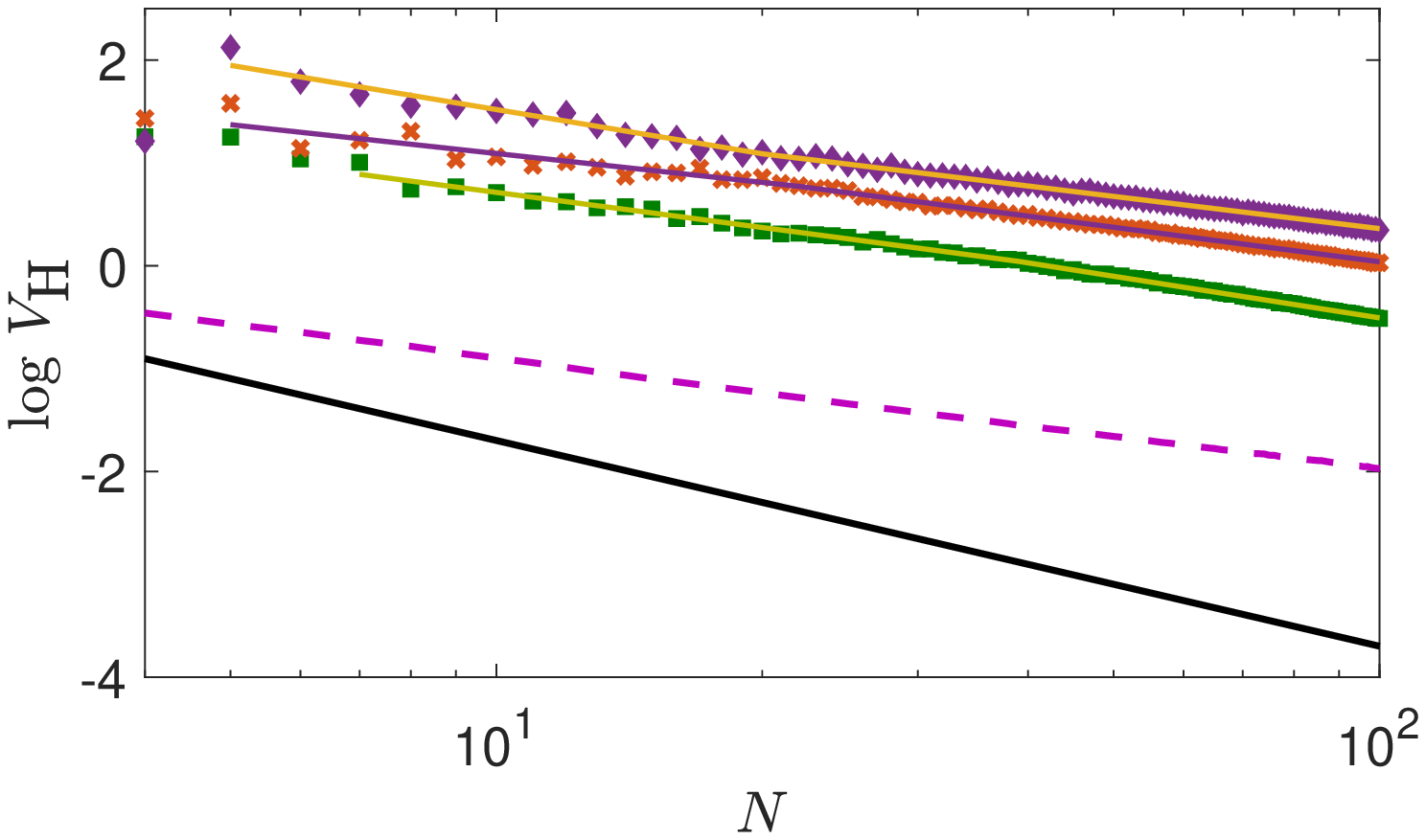}}
	%(c)skew-normal-distribution noise\\
	\subfloat[log-normal-distribution noise]{\includegraphics[width=0.45\textwidth]{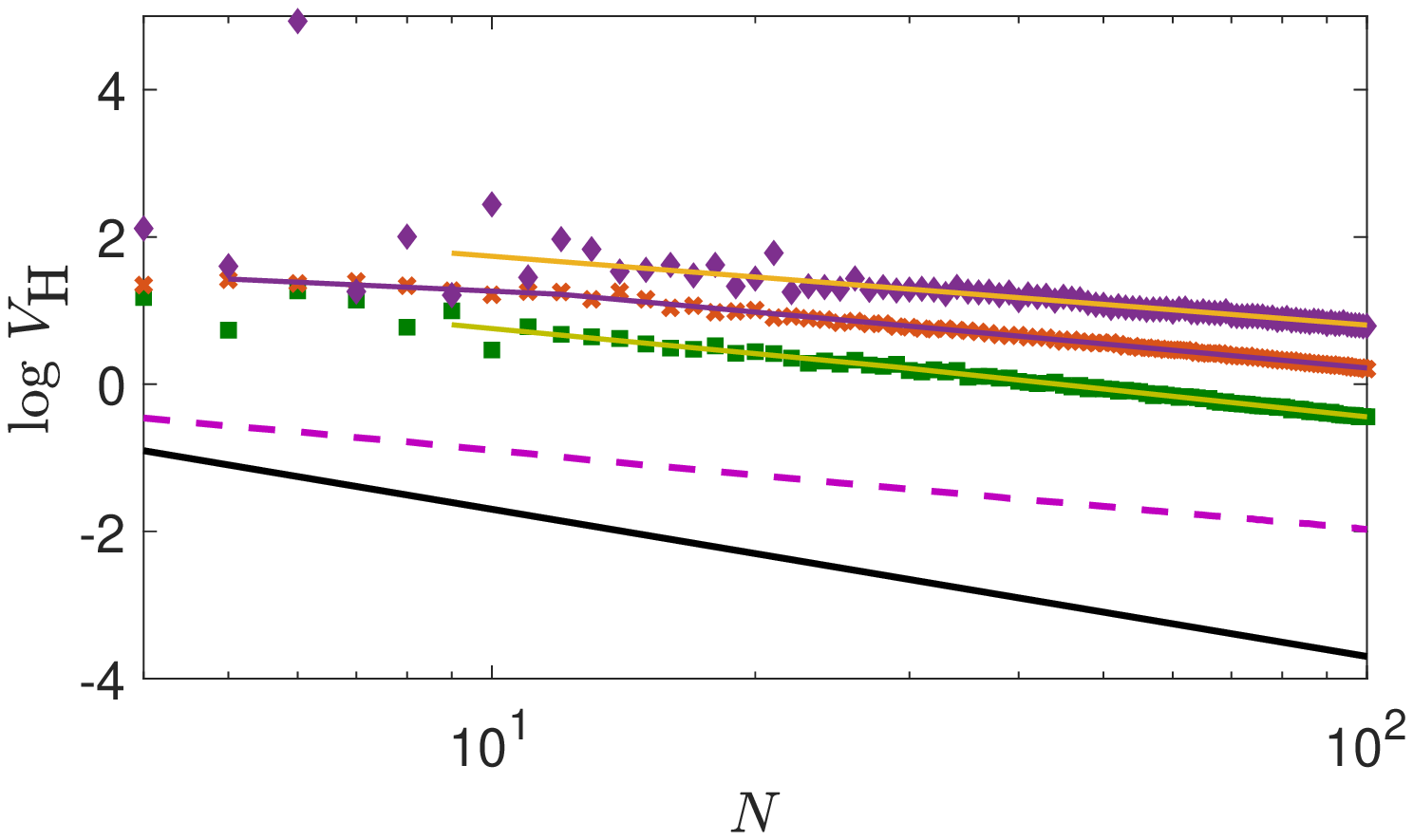}}
	%(d)log-normal-distribution noise\\
	\caption{\label{fig:noise data BF}Logarithmic plots of Holevo variance from simulation of adaptive phase estimations that use Bayesian feedback method. The simulation includes from the four noise models, namely, (a) normal-distribution noise, (b) random telegraph noise (c), skew-normal-distribution noise, and (d) log-normal-distribution noise.
		The plot for the normal-distribution noise also includes the data from the noiseless simulation (brown side-facing triangle) and its linear fit (green solid).
		The blue circles are data when $V=1$, the red triangles when $V=2$, the green squares when $V=3$, the brown plus when $V=4$, the brown crosses when $V=5$ , and the purple diamonds when $V=7$. 
		The lines shown are the piecewise~linear fit of the data whose scaling is reported.
		The solid black line if the HL and the dashed purple line in the SQL generated from noiseless adaptive phase estimation.}	
\end{figure*}

Log-log plots of $V_\text{H}$ as a function of $N$,
shown in Fig.~\ref{fig:noise data BF},
are computed from simulations of adaptive phase estimation controlled by Bayesian feedback.
Figures~\ref{fig:noise data BF}(a--d) present~$V_\text{H}$ in the presence of normal-distribution noise, random telegraph noise, skew-normal noise, and log-normal noise respectively.

Similar to Fig.~\ref{fig:noise data RL}, the  trend of~$V_\text{H}$ vs~$N$ in Fig.~\ref{fig:noise data BF} shows that the power-law relationship also breaks into parts. Instead of a bump,
$V_\text{H}$ from Bayesian feedback exhibits noise for low~$N$.
For this reason, the model-selection procedure~\ref{subsubsec:modelselection} favours model with linear interpolation in the first segment.
The subsequent segment appears straight in the log-log plots, although some, such as the~$V=7$ in Fig.~\ref{fig:noise data BF}(c), shows a break into two linear segments.

The intercepts of the $V_\text{H}$ vs $N$ plots increase with the increase of $V$, and the observation of the changes are similar to when reinforcement-learning policies are used (\S\ref{subsubsec:RLvariance}).
The asymmetric noise show slow increase in intercept when compared to symmetric noise,
and the rate of change depends on the noise model.

\subsection{Power-law scaling}
\label{subsec:powerlawscaling}
In this subsection, we present values of~$\wp$, summarized in Table.~\ref{tab:robustness}, that are estimated by fitting~$V_\text{H}$ plots in~\S\ref{subsec:variancen}.
\begin{table}	\caption{\label{tab:robustness}Power-law scaling from adaptive phase estimation in noisy condition
	using reinforcement-learning policies $\wp_\text{R}$ and Bayesian feedback~$\wp_\text{B}$. 
	}
	%\begin{ruledtabular}
	\begin{tabular}{|l|c|c|c|c|c|c|}
		\hline
		\textbf{} & $V$ &~$\gamma$ &$2\wp_{\text{R}}$ & $\overline{R^2}_{\text{R}}$ & $2\wp_{\text{B}}$&$\overline{R^2}_{\text{B}}$\\
		\hline
		SQL & & & 1 & & 1 &\\
		HL & & & 2 & & 2&\\
		No noise & & & 1.459&0.9998
		&1.957&0.9993
		\\
		\hline
		& 1 & 0 &$1.302$&0.9999&$1.512$&0.9985
		\\
		Normal & 2 & 0 &$1.267$&0.9999& -- &--\\
		& 3 & 0 &$0.954$&0.9992
		&$1.190$&0.9997
		\\
		& 4 & 0 & -- &--&$1.004$&0.9948
		\\
		\hline
		& 1 & 0 &$1.266$&0.9999
		&$1.526$&0.9991\\
		Random telegraph & 2 & 0 &$1.186$&0.9997
		&$1.277$&0.9967\\
		&3 & 0 &$0.935$&0.9993
		&$0.919$&0.9892
		\\
		\hline
		& 1 & 0.8509 &$1.296$&0.9999
		&--&--\\
		Skew-normal & 3 & 0.8509&$1.246$&0.9999
		&$1.343$&0.9987
		\\
		& 5 & 0.8509&$1.118$&0.9998
		&$1.116$&0.9927\\
		& 7& 0.8509&$1.039$&0.9996
		&$1.041$&0.9964\\
		\hline
		& 1 &0.8509&$1.290$&0.9999
		&--&--\\
		Log-normal& 3 &0.8509&$1.217$&0.9998
		&$1.258$&0.9919\\
		&5 &0.8509&$1.058$&0.9997
		&$1.086$&0.9961\\
		&7 &0.8509&$0.981$&0.9994
		&$0.9209$&0.7965
		\\
		\hline
	\end{tabular}
	%\end{ruledtabular}
\end{table}
These~$\wp$'s are from the last segment of the selected piecewise linear models (\S\ref{subsub:piecewise models}), which changes with the increase in $V$. 
We also include $\overline{R^2}$ (\ref{eq:adjusted R-squared}) to show the goodness of fit.

The power-law scaling for reinforcement-learning policy $\wp_\text{R}$ shows a decrease as the noise level $V$ increases, starting from the noiseless phase estimation at $2\wp_\text{R}$=1.459.
The reinforcement-learning policies fail to deliver $\wp_\text{R}>1/2$ when $V=3$ for the symmetric noise distributions. This limit increases with asymmetric noise models to $V=7$ in log-normal noise. The skew-normal noise only shows a scaling at approaches the SQL but does not breach it at all.

Similar trends are observed for the Bayesian feedback.
The scaling $\wp_\text{B}$ from noiseless interferometer closely approximates the HL at $2\wp_\text{B}=1.957$
and approaches SQL when $V=4$
for normal-distribution noise.
This limit drops to $V=3$ when random telegraph noise is included.
This limit also appears at $V=7$ for log-normal noise, whereas the same noise-level only lead to $\wp_\text{B}$ approaching SQL when skew-normal noise is present.
These trend, aside from the case of normal-distributed noise, is the same as the trend for the reinforcement-learning policies.

The goodness-of-fit for these fits are reported in term of  $\overline{R^2}$, where $\overline{R^2}=1$ indicates a perfect fit.
The values of the goodness $\overline{R^2}_\text{R}>0.999$ for $V_\text{H}$ delivered by reinforcement-learning policies and $\overline{R^2}_\text{B}>0.99$ for Bayesian feedback except for when a log-normal noise of $V=7$ is present.
Overall, the models chosen using the method in \S\ref{subsubsec:modelselection} provide good fits to the data and the reinforcement-learning policies always deliver fits with $\overline{R^2}_\text{R}>\overline{R^2}_\text{B}$.

\subsection{Bounds on time and space costs}
\label{subsec:bounds}
The result from the calculation of space and time complexities for both generating and implementing the reinforcement-learning policies and the Bayesian feedback is shown in Table~\ref{tab:complexity}.
\begin{table}
	\caption{\label{tab:complexity}Upper bound in policy space and time cost of the policy from reinforcement-learning algorithm~(RL) and the Bayesian feedback~(BF).}
	%\begin{ruledtabular}
	\begin{tabular}{|c|c|c|}
		\hline
		Complexity & RL & BF\\
		\hline
		Design time & $O(N^6)$ & --\\
		\hline
		Policy space & $O(N)$ & $O(N^2)$\\		
		Implementation time & $O(N)$ & $O(N^3)$\\
		\hline
	\end{tabular}
	%\end{ruledtabular}
\end{table}
Here we compare how these results.

The time complexity for generating policies, called here the design time, is of high polynomial when reinforcement-learning algorithm is used. 
Bayesian feedback, which is designed through an analytical process, incurs no time cost for the design .
When the implementation time is compared, the time complexity of Bayesian feedback is two order above the reinforcement-learning policies.
The space complexity, which shows the memory used to store feedback information, is also larger for  Bayesian feedback than are reinforcement-learning policies by an polynomial degree,
specifically, going from linear scaling in the reinforcement-learning policy
to quadratic for Bayesian feedback.

\section{Discussion\label{sec:discussion}}
%up to you but, this explanation of the discussion is rather redundant and might not be necessary%
In this section, we discuss the robustness and its threshold for AQEM policies and possible reasons behind the high level of noise-resistant. 
We explain the difference between the $\wp$ attained by reinforcement-learning policies and Bayesian feedback
using the space complexity of the policies.
We opt for policies and methods for generating policies based on the resource complexity of the generated policies.

\subsection{Robustness of AQEM policies}
\label{subsec:robustnessAQEMpolicies}
In this subsection, we discuss robustness for the AQEM policies and the robustness threshold based on~$\wp$ in Table~\ref{tab:robustness}.
Here we discuss the robustness threshold and propose explanations for this high threshold.

Both the reinforcement-learning policies and the Bayesian feedback are able to deliver $\wp>1/2$ for all four noise models until $V=3$ when the scaling obtained in the present of random telegraph noise fails to exceed the SQL. For noise models that are asymmetric, quantum-enhanced precision is observed up to $V=7$ at the skewness of $\gamma=0.8509$. This high level of robustness is unexpected, and especially so for the Bayesian feedback as the dynamic of the interferometer no longer matches the noiseless assumption.

One possible reason behind this high robustness threshold is the sine state (\ref{eq:sinestate}), 
which is already known to be robust against loss~\cite{HS11a}. 
The structure of the sine state may also contribute to robustness against phase noise as well,
although AQEM policies also play a role in the robustness of the AQEM scheme.

The effect of the feedback policy is highlighted by the threshold for normal-distribution noise,
where the threshold for reinforcement-learning policies is at $V=3$ as opposed to the Bayesian feedback at $V=4$.
This result, however, does not imply that robustness against noise of unknown distribution is improved
as the thresholds for all other noise models are the same for both the reinforcement-learning policies and the Bayesian feedback.

\subsection{Space cost and power-law scalings}
Table~\ref{tab:robustness} shows that the power-law scaling delivered
by Bayesian feedback is consistently superior to those delivered by the reinforcement-learning policies before the robustness threshold is reached. 
Here we use the space complexity of the policies in Table~\ref{tab:complexity} to explain the reason behind this difference.

Table~\ref{tab:complexity} shows that the Bayesian feedback has a space cost that scales a polynomial degree higher
than for the reinforcement-learning policies,
which indicates that the Bayesian feedback utilizes more information and hence is more complex than the reinforcement-learning policies.
By using a quantum-state model,
Bayesian feedback effectively uses the history of measurement outcomes $x_1x_2\cdots x_m$ to determine $\Phi_{m+1}$ instead of the current outcome $x_m$, which is the approach used by the reinforcement-learning policies in (\ref{eq:rl update}).
As such,
reinforcement-learning policies are restricted
by the generalized-logarithm-search strategy (\S\ref{subsubsec:policygen}) and so cannot deliver a value of~$\wp_{\text{R}}$
that  approaches the HL.
Improvement of reinforcement-learning policies can be done by changing the update rule so that the policy uses a part of the measurement history.

\subsection{Choosing an AQEM policy}
In this subsection, we explain how the space and implementation time complexity (\S\ref{subsubsec:policycomplexity}) can be used to decide between competing policies and method of generating the policies.
In particular, we discuss choosing between the reinforcement-learning policies and the Bayesian feedback. 

The consideration of the space and time complexity of the policies comes after ascertaining that the candidate policies are able to deliver the target performance. 
In the case of AQEM, the target is to attain $\wp>1/2$, which both the reinforcement-learning policies and the Bayesian feedback are able to deliver.
Both methods also have the same robustness threshold against phase noise of unknown distribution.
Based on these comparison, both policies appears equally suitable.

When comparing space and implementation complexity (\S\ref{subsec:bounds}),
reinforcement-learning policies shows an advantage as the scaling of both costs are linearly bound, whereas the Bayesian feedback is quadratic in space complexity and cubic in time complexity. For this reason, we favour the reinforcement-learning policies for robust adaptive phase estimation.

We do not consider the design complexity, which is used in the generation of the policies, 
as the scaling of the cost can be improved by parallelizing the training. 
This cost may be of interest when the learning occurs in a physical setup where parallelizing is not possible and one shot of the experiment is expensive.
In this case, there maybe an upper bound in number of experiments and hence time that can be invested in training a policy, and the design complexity can be used to infer the method that generate a policy within this bound.

\section{Conclusion\label{sec:conclusion}}
We have tested AQEM policy robustness 
based on Bayesian feedback and reinforcement learning and compared the resource complexities 
for implementing the policies.
We find that both the reinforcement-learning policies and Bayesian feedback are robust against phase noise up to noise level of~$V=3$.
Although the imprecision scaling delivered
by Bayesian feedback is superior to the policies from reinforcement learning, the latter policies use less resource than does Bayesian feedback 
with respect to both space and time complexity.

Although we develop a robustness test and method of comparing policies based on resource complexity with AQEM procedures in mind, these ideas can be applied to other cases in QEM and quantum control.
Robustness is a desirable property of any QEM schemes
and the test presented here can be adapted to quantify robustness of non-adaptive procedures and investigate the role of the input state to the robustness of QEM generally.
Quantifying the resource used by control policies can be used to show efficacy of the policies not only in AQEM but in other quantum control tasks, and the comparison of the resource complexity can be used to select a policy that is most efficient in accomplishing a task.

\begin{acknowledgments}
The computational work is enabled by the support of WestGrid (www.westgrid.ca) through Compute Canada Calcul Canada (www.computecanada.ca) and by 
NSERC and AITF.
\end{acknowledgments}
\bibliography{robust}

%merlin.mbs apsrev4-1.bst 2010-07-25 4.21a (PWD, AO, DPC) hacked
%Control: key (0)
%Control: author (8) initials jnrlst
%Control: editor formatted (1) identically to author
%Control: production of article title (-1) disabled
%Control: page (0) single
%Control: year (1) truncated
%Control: production of eprint (0) enabled
\begin{thebibliography}{90}%
\makeatletter
\providecommand \@ifxundefined [1]{%
 \@ifx{#1\undefined}
}%
\providecommand \@ifnum [1]{%
 \ifnum #1\expandafter \@firstoftwo
 \else \expandafter \@secondoftwo
 \fi
}%
\providecommand \@ifx [1]{%
 \ifx #1\expandafter \@firstoftwo
 \else \expandafter \@secondoftwo
 \fi
}%
\providecommand \natexlab [1]{#1}%
\providecommand \enquote  [1]{``#1''}%
\providecommand \bibnamefont  [1]{#1}%
\providecommand \bibfnamefont [1]{#1}%
\providecommand \citenamefont [1]{#1}%
\providecommand \href@noop [0]{\@secondoftwo}%
\providecommand \href [0]{\begingroup \@sanitize@url \@href}%
\providecommand \@href[1]{\@@startlink{#1}\@@href}%
\providecommand \@@href[1]{\endgroup#1\@@endlink}%
\providecommand \@sanitize@url [0]{\catcode `\\12\catcode `\$12\catcode
  `\&12\catcode `\#12\catcode `\^12\catcode `\_12\catcode `\%12\relax}%
\providecommand \@@startlink[1]{}%
\providecommand \@@endlink[0]{}%
\providecommand \url  [0]{\begingroup\@sanitize@url \@url }%
\providecommand \@url [1]{\endgroup\@href {#1}{\urlprefix }}%
\providecommand \urlprefix  [0]{URL }%
\providecommand \Eprint [0]{\href }%
\providecommand \doibase [0]{http://dx.doi.org/}%
\providecommand \selectlanguage [0]{\@gobble}%
\providecommand \bibinfo  [0]{\@secondoftwo}%
\providecommand \bibfield  [0]{\@secondoftwo}%
\providecommand \translation [1]{[#1]}%
\providecommand \BibitemOpen [0]{}%
\providecommand \bibitemStop [0]{}%
\providecommand \bibitemNoStop [0]{.\EOS\space}%
\providecommand \EOS [0]{\spacefactor3000\relax}%
\providecommand \BibitemShut  [1]{\csname bibitem#1\endcsname}%
\let\auto@bib@innerbib\@empty
%</preamble>
\bibitem [{\citenamefont {Giovannetti}\ \emph {et~al.}(2004)\citenamefont
  {Giovannetti}, \citenamefont {Lloyd},\ and\ \citenamefont {Maccone}}]{GLM04}%
  \BibitemOpen
  \bibfield  {author} {\bibinfo {author} {\bibfnamefont {V.}~\bibnamefont
  {Giovannetti}}, \bibinfo {author} {\bibfnamefont {S.}~\bibnamefont {Lloyd}},
  \ and\ \bibinfo {author} {\bibfnamefont {L.}~\bibnamefont {Maccone}},\ }\href
  {\doibase 10.1126/science.1104149} {\bibfield  {journal} {\bibinfo  {journal}
  {Science}\ }\textbf {\bibinfo {volume} {306}},\ \bibinfo {pages} {1330}
  (\bibinfo {year} {2004})}\BibitemShut {NoStop}%
\bibitem [{\citenamefont {Giovannetti}\ \emph {et~al.}(2011)\citenamefont
  {Giovannetti}, \citenamefont {Lloyd},\ and\ \citenamefont {Maccone}}]{GLM11}%
  \BibitemOpen
  \bibfield  {author} {\bibinfo {author} {\bibfnamefont {V.}~\bibnamefont
  {Giovannetti}}, \bibinfo {author} {\bibfnamefont {S.}~\bibnamefont {Lloyd}},
  \ and\ \bibinfo {author} {\bibfnamefont {L.}~\bibnamefont {Maccone}},\ }\href
  {\doibase 10.1038/nphoton.2011.35} {\bibfield  {journal} {\bibinfo  {journal}
  {Nat. Photon.}\ }\textbf {\bibinfo {volume} {5}},\ \bibinfo {pages} {222}
  (\bibinfo {year} {2011})}\BibitemShut {NoStop}%
\bibitem [{\citenamefont {Braunstein}\ \emph {et~al.}(1996)\citenamefont
  {Braunstein}, \citenamefont {Caves},\ and\ \citenamefont {Milburn}}]{BCM96}%
  \BibitemOpen
  \bibfield  {author} {\bibinfo {author} {\bibfnamefont {S.~L.}\ \bibnamefont
  {Braunstein}}, \bibinfo {author} {\bibfnamefont {C.~M.}\ \bibnamefont
  {Caves}}, \ and\ \bibinfo {author} {\bibfnamefont {G.~J.}\ \bibnamefont
  {Milburn}},\ }\href {\doibase 10.1006/aphy.1996.0040} {\bibfield  {journal}
  {\bibinfo  {journal} {Ann. Phys. (N. Y.)}\ }\textbf {\bibinfo {volume}
  {247}},\ \bibinfo {pages} {135} (\bibinfo {year} {1996})}\BibitemShut
  {NoStop}%
\bibitem [{\citenamefont {T\'{o}th}\ and\ \citenamefont
  {Apellaniz}(2014)}]{TA14}%
  \BibitemOpen
  \bibfield  {author} {\bibinfo {author} {\bibfnamefont {G.}~\bibnamefont
  {T\'{o}th}}\ and\ \bibinfo {author} {\bibfnamefont {I.}~\bibnamefont
  {Apellaniz}},\ }\href {\doibase 10.1088/1751-8113/47/42/424006} {\bibfield
  {journal} {\bibinfo  {journal} {J. Phys. A: Math. Theor.}\ }\textbf {\bibinfo
  {volume} {47}},\ \bibinfo {pages} {424006} (\bibinfo {year}
  {2014})}\BibitemShut {NoStop}%
\bibitem [{\citenamefont {Vrajitoru}\ and\ \citenamefont
  {Knight}(2014{\natexlab{a}})}]{VK14_ch1}%
  \BibitemOpen
  \bibfield  {author} {\bibinfo {author} {\bibfnamefont {D.}~\bibnamefont
  {Vrajitoru}}\ and\ \bibinfo {author} {\bibfnamefont {W.}~\bibnamefont
  {Knight}},\ }\enquote {\bibinfo {title} {Introduction},}\ in\ \href {\doibase
  10.1007/978-3-319-09888-3_1} {\emph {\bibinfo {booktitle} {Practical Analysis
  of Algorithms}}}\ (\bibinfo  {publisher} {Springer},\ \bibinfo {address}
  {Cham, Switzerland},\ \bibinfo {year} {2014})\ Chap.~\bibinfo {chapter} {1},
  pp.\ \bibinfo {pages} {1--7}\BibitemShut {NoStop}%
\bibitem [{\citenamefont {Hollenhorst}(1979)}]{Hol79}%
  \BibitemOpen
  \bibfield  {author} {\bibinfo {author} {\bibfnamefont {J.~N.}\ \bibnamefont
  {Hollenhorst}},\ }\href {\doibase 10.1103/PhysRevD.19.1669} {\bibfield
  {journal} {\bibinfo  {journal} {Phys. Rev. D}\ }\textbf {\bibinfo {volume}
  {19}},\ \bibinfo {pages} {1669} (\bibinfo {year} {1979})}\BibitemShut
  {NoStop}%
\bibitem [{\citenamefont {Caves}\ \emph {et~al.}(1980)\citenamefont {Caves},
  \citenamefont {Thorne}, \citenamefont {Drever}, \citenamefont {Sandberg},\
  and\ \citenamefont {Zimmermann}}]{CTD+80}%
  \BibitemOpen
  \bibfield  {author} {\bibinfo {author} {\bibfnamefont {C.~M.}\ \bibnamefont
  {Caves}}, \bibinfo {author} {\bibfnamefont {K.~S.}\ \bibnamefont {Thorne}},
  \bibinfo {author} {\bibfnamefont {R.~W.~P.}\ \bibnamefont {Drever}}, \bibinfo
  {author} {\bibfnamefont {V.~D.}\ \bibnamefont {Sandberg}}, \ and\ \bibinfo
  {author} {\bibfnamefont {M.}~\bibnamefont {Zimmermann}},\ }\href {\doibase
  10.1103/RevModPhys.52.341} {\bibfield  {journal} {\bibinfo  {journal} {Rev.
  Mod. Phys.}\ }\textbf {\bibinfo {volume} {52}},\ \bibinfo {pages} {341}
  (\bibinfo {year} {1980})}\BibitemShut {NoStop}%
\bibitem [{\citenamefont {Caves}(1981)}]{Cav81}%
  \BibitemOpen
  \bibfield  {author} {\bibinfo {author} {\bibfnamefont {C.~M.}\ \bibnamefont
  {Caves}},\ }\href {\doibase 10.1103/PhysRevD.23.1693} {\bibfield  {journal}
  {\bibinfo  {journal} {Phys. Rev. D}\ }\textbf {\bibinfo {volume} {23}},\
  \bibinfo {pages} {1693} (\bibinfo {year} {1981})}\BibitemShut {NoStop}%
\bibitem [{\citenamefont {Bollinger}\ \emph {et~al.}(1996)\citenamefont
  {Bollinger}, \citenamefont {Itano}, \citenamefont {Wineland},\ and\
  \citenamefont {Heinzen}}]{BIWH96}%
  \BibitemOpen
  \bibfield  {author} {\bibinfo {author} {\bibfnamefont {J.~J.~.}\ \bibnamefont
  {Bollinger}}, \bibinfo {author} {\bibfnamefont {W.~M.}\ \bibnamefont
  {Itano}}, \bibinfo {author} {\bibfnamefont {D.~J.}\ \bibnamefont {Wineland}},
  \ and\ \bibinfo {author} {\bibfnamefont {D.~J.}\ \bibnamefont {Heinzen}},\
  }\href {\doibase 10.1103/PhysRevA.54.R4649} {\bibfield  {journal} {\bibinfo
  {journal} {Phys. Rev. A}\ }\textbf {\bibinfo {volume} {54}},\ \bibinfo
  {pages} {R4649} (\bibinfo {year} {1996})}\BibitemShut {NoStop}%
\bibitem [{\citenamefont {Borregaard}\ and\ \citenamefont
  {S\o{}rensen}(2013)}]{BS13}%
  \BibitemOpen
  \bibfield  {author} {\bibinfo {author} {\bibfnamefont {J.}~\bibnamefont
  {Borregaard}}\ and\ \bibinfo {author} {\bibfnamefont {A.~S.}\ \bibnamefont
  {S\o{}rensen}},\ }\href {\doibase 10.1103/PhysRevLett.111.090801} {\bibfield
  {journal} {\bibinfo  {journal} {Phys. Rev. Lett.}\ }\textbf {\bibinfo
  {volume} {111}},\ \bibinfo {pages} {090801} (\bibinfo {year}
  {2013})}\BibitemShut {NoStop}%
\bibitem [{\citenamefont {Danilin}\ \emph {et~al.}(2018)\citenamefont
  {Danilin}, \citenamefont {Lebedev}, \citenamefont {Veps{\"a}l{\"a}inen},
  \citenamefont {Lesovik}, \citenamefont {Blatter},\ and\ \citenamefont
  {Paraoanu}}]{DLV+18}%
  \BibitemOpen
  \bibfield  {author} {\bibinfo {author} {\bibfnamefont {S.}~\bibnamefont
  {Danilin}}, \bibinfo {author} {\bibfnamefont {A.~V.}\ \bibnamefont
  {Lebedev}}, \bibinfo {author} {\bibfnamefont {A.}~\bibnamefont
  {Veps{\"a}l{\"a}inen}}, \bibinfo {author} {\bibfnamefont {G.~B.}\
  \bibnamefont {Lesovik}}, \bibinfo {author} {\bibfnamefont {G.}~\bibnamefont
  {Blatter}}, \ and\ \bibinfo {author} {\bibfnamefont {G.~S.}\ \bibnamefont
  {Paraoanu}},\ }\href {\doibase 10.1038/s41534-018-0078-y} {\bibfield
  {journal} {\bibinfo  {journal} {npj Quantum Inf.}\ }\textbf {\bibinfo
  {volume} {4}},\ \bibinfo {pages} {29} (\bibinfo {year} {2018})}\BibitemShut
  {NoStop}%
\bibitem [{\citenamefont {Zhang}\ and\ \citenamefont {Duan}(2014)}]{ZD14}%
  \BibitemOpen
  \bibfield  {author} {\bibinfo {author} {\bibfnamefont {Z.}~\bibnamefont
  {Zhang}}\ and\ \bibinfo {author} {\bibfnamefont {L.~M.}\ \bibnamefont
  {Duan}},\ }\href@noop {} {\bibfield  {journal} {\bibinfo  {journal} {New J.
  Phys.}\ }\textbf {\bibinfo {volume} {16}},\ \bibinfo {pages} {103037}
  (\bibinfo {year} {2014})}\BibitemShut {NoStop}%
\bibitem [{\citenamefont {Matthews}\ \emph {et~al.}(2016)\citenamefont
  {Matthews}, \citenamefont {Zhou}, \citenamefont {Cable}, \citenamefont
  {Shadbolt}, \citenamefont {Saunders}, \citenamefont {Durkin}, \citenamefont
  {Pryde},\ and\ \citenamefont {O'Brien}}]{MZC+16}%
  \BibitemOpen
  \bibfield  {author} {\bibinfo {author} {\bibfnamefont {J.~C.~F.}\
  \bibnamefont {Matthews}}, \bibinfo {author} {\bibfnamefont {X.-Q.}\
  \bibnamefont {Zhou}}, \bibinfo {author} {\bibfnamefont {H.}~\bibnamefont
  {Cable}}, \bibinfo {author} {\bibfnamefont {P.~J.}\ \bibnamefont {Shadbolt}},
  \bibinfo {author} {\bibfnamefont {D.~J.}\ \bibnamefont {Saunders}}, \bibinfo
  {author} {\bibfnamefont {G.~A.}\ \bibnamefont {Durkin}}, \bibinfo {author}
  {\bibfnamefont {G.~J.}\ \bibnamefont {Pryde}}, \ and\ \bibinfo {author}
  {\bibfnamefont {J.~L.}\ \bibnamefont {O'Brien}},\ }\href {\doibase
  10.1038/npjqi.2016.23} {\bibfield  {journal} {\bibinfo  {journal} {Npj
  Quantum Inf.}\ }\textbf {\bibinfo {volume} {2}},\ \bibinfo {pages} {16023}
  (\bibinfo {year} {2016})}\BibitemShut {NoStop}%
\bibitem [{\citenamefont {Wiseman}(1995)}]{Wis95}%
  \BibitemOpen
  \bibfield  {author} {\bibinfo {author} {\bibfnamefont {H.~M.}\ \bibnamefont
  {Wiseman}},\ }\href {\doibase 10.1103/PhysRevLett.75.4587} {\bibfield
  {journal} {\bibinfo  {journal} {Phys. Rev. Lett.}\ }\textbf {\bibinfo
  {volume} {75}},\ \bibinfo {pages} {4587} (\bibinfo {year}
  {1995})}\BibitemShut {NoStop}%
\bibitem [{\citenamefont {Armen}\ \emph {et~al.}(2002)\citenamefont {Armen},
  \citenamefont {Au}, \citenamefont {Stockton}, \citenamefont {Doherty},\ and\
  \citenamefont {Mabuchi}}]{AAS+02}%
  \BibitemOpen
  \bibfield  {author} {\bibinfo {author} {\bibfnamefont {M.~A.}\ \bibnamefont
  {Armen}}, \bibinfo {author} {\bibfnamefont {J.~K.}\ \bibnamefont {Au}},
  \bibinfo {author} {\bibfnamefont {J.~K.}\ \bibnamefont {Stockton}}, \bibinfo
  {author} {\bibfnamefont {A.~C.}\ \bibnamefont {Doherty}}, \ and\ \bibinfo
  {author} {\bibfnamefont {H.}~\bibnamefont {Mabuchi}},\ }\href {\doibase
  10.1103/PhysRevLett.89.133602} {\bibfield  {journal} {\bibinfo  {journal}
  {Phys. Rev. Lett.}\ }\textbf {\bibinfo {volume} {89}},\ \bibinfo {pages}
  {133602} (\bibinfo {year} {2002})}\BibitemShut {NoStop}%
\bibitem [{\citenamefont {Berry}\ and\ \citenamefont {Wiseman}(2000)}]{BW00}%
  \BibitemOpen
  \bibfield  {author} {\bibinfo {author} {\bibfnamefont {D.~W.}\ \bibnamefont
  {Berry}}\ and\ \bibinfo {author} {\bibfnamefont {H.~M.}\ \bibnamefont
  {Wiseman}},\ }\href {\doibase 10.1103/PhysRevLett.85.5098} {\bibfield
  {journal} {\bibinfo  {journal} {Phys. Rev. Lett.}\ }\textbf {\bibinfo
  {volume} {85}},\ \bibinfo {pages} {5098} (\bibinfo {year}
  {2000})}\BibitemShut {NoStop}%
\bibitem [{\citenamefont {Rosolia}\ \emph {et~al.}(2018)\citenamefont
  {Rosolia}, \citenamefont {Zhang},\ and\ \citenamefont {Borrelli}}]{RZB18}%
  \BibitemOpen
  \bibfield  {author} {\bibinfo {author} {\bibfnamefont {U.}~\bibnamefont
  {Rosolia}}, \bibinfo {author} {\bibfnamefont {X.}~\bibnamefont {Zhang}}, \
  and\ \bibinfo {author} {\bibfnamefont {F.}~\bibnamefont {Borrelli}},\ }\href
  {\doibase 10.1146/annurev-control-060117-105215} {\bibfield  {journal}
  {\bibinfo  {journal} {Annual Review of Control, Robotics, and Autonomous
  Systems}\ }\textbf {\bibinfo {volume} {1}},\ \bibinfo {pages} {259} (\bibinfo
  {year} {2018})}\BibitemShut {NoStop}%
\bibitem [{\citenamefont {Roy}\ \emph {et~al.}(2015)\citenamefont {Roy},
  \citenamefont {Petersen},\ and\ \citenamefont {Huntington}}]{RPH15}%
  \BibitemOpen
  \bibfield  {author} {\bibinfo {author} {\bibfnamefont {S.}~\bibnamefont
  {Roy}}, \bibinfo {author} {\bibfnamefont {I.~R.}\ \bibnamefont {Petersen}}, \
  and\ \bibinfo {author} {\bibfnamefont {E.~H.}\ \bibnamefont {Huntington}},\
  }\href {\doibase 10.1088/1367-2630/17/6/063020} {\bibfield  {journal}
  {\bibinfo  {journal} {New J. Phys.}\ }\textbf {\bibinfo {volume} {17}},\
  \bibinfo {pages} {063020} (\bibinfo {year} {2015})}\BibitemShut {NoStop}%
\bibitem [{\citenamefont {Wiseman}\ and\ \citenamefont {Killip}(1997)}]{WK97}%
  \BibitemOpen
  \bibfield  {author} {\bibinfo {author} {\bibfnamefont {H.~M.}\ \bibnamefont
  {Wiseman}}\ and\ \bibinfo {author} {\bibfnamefont {R.}~\bibnamefont
  {Killip}},\ }\href {\doibase 10.1103/PhysRevA.56.944} {\bibfield  {journal}
  {\bibinfo  {journal} {Phys. Rev. A}\ }\textbf {\bibinfo {volume} {56}},\
  \bibinfo {pages} {944} (\bibinfo {year} {1997})}\BibitemShut {NoStop}%
\bibitem [{\citenamefont {Wiseman}\ and\ \citenamefont {Killip}(1998)}]{WK98}%
  \BibitemOpen
  \bibfield  {author} {\bibinfo {author} {\bibfnamefont {H.~M.}\ \bibnamefont
  {Wiseman}}\ and\ \bibinfo {author} {\bibfnamefont {R.}~\bibnamefont
  {Killip}},\ }\href {\doibase 10.1103/PhysRevA.57.2169} {\bibfield  {journal}
  {\bibinfo  {journal} {Phys. Rev. A}\ }\textbf {\bibinfo {volume} {57}},\
  \bibinfo {pages} {2169} (\bibinfo {year} {1998})}\BibitemShut {NoStop}%
\bibitem [{\citenamefont {Hentschel}\ and\ \citenamefont
  {Sanders}(2010)}]{HS10}%
  \BibitemOpen
  \bibfield  {author} {\bibinfo {author} {\bibfnamefont {A.}~\bibnamefont
  {Hentschel}}\ and\ \bibinfo {author} {\bibfnamefont {B.~C.}\ \bibnamefont
  {Sanders}},\ }\href {\doibase 10.1103/PhysRevLett.104.063603} {\bibfield
  {journal} {\bibinfo  {journal} {Phys. Rev. Lett.}\ }\textbf {\bibinfo
  {volume} {104}},\ \bibinfo {pages} {063603} (\bibinfo {year}
  {2010})}\BibitemShut {NoStop}%
\bibitem [{\citenamefont {Lovett}\ \emph {et~al.}(2013)\citenamefont {Lovett},
  \citenamefont {Crosnier}, \citenamefont {Perarnau-Llobet},\ and\
  \citenamefont {Sanders}}]{LCPS13}%
  \BibitemOpen
  \bibfield  {author} {\bibinfo {author} {\bibfnamefont {N.~B.}\ \bibnamefont
  {Lovett}}, \bibinfo {author} {\bibfnamefont {C.}~\bibnamefont {Crosnier}},
  \bibinfo {author} {\bibfnamefont {M.}~\bibnamefont {Perarnau-Llobet}}, \ and\
  \bibinfo {author} {\bibfnamefont {B.~C.}\ \bibnamefont {Sanders}},\ }\href
  {\doibase 10.1103/PhysRevLett.110.220501} {\bibfield  {journal} {\bibinfo
  {journal} {Phys. Rev. Lett.}\ }\textbf {\bibinfo {volume} {110}},\ \bibinfo
  {pages} {220501} (\bibinfo {year} {2013})}\BibitemShut {NoStop}%
\bibitem [{\citenamefont {Palittapongarnpim}\ \emph
  {et~al.}(2017{\natexlab{a}})\citenamefont {Palittapongarnpim}, \citenamefont
  {Wittek}, \citenamefont {Zahedinejad}, \citenamefont {Vedaie},\ and\
  \citenamefont {Sanders}}]{PWZ+17}%
  \BibitemOpen
  \bibfield  {author} {\bibinfo {author} {\bibfnamefont {P.}~\bibnamefont
  {Palittapongarnpim}}, \bibinfo {author} {\bibfnamefont {P.}~\bibnamefont
  {Wittek}}, \bibinfo {author} {\bibfnamefont {E.}~\bibnamefont {Zahedinejad}},
  \bibinfo {author} {\bibfnamefont {S.}~\bibnamefont {Vedaie}}, \ and\ \bibinfo
  {author} {\bibfnamefont {B.~C.}\ \bibnamefont {Sanders}},\ }\href {\doibase
  10.1016/j.neucom.2016.12.087} {\bibfield  {journal} {\bibinfo  {journal}
  {Neurocomputing}\ }\textbf {\bibinfo {volume} {268}},\ \bibinfo {pages} {116}
  (\bibinfo {year} {2017}{\natexlab{a}})}\BibitemShut {NoStop}%
\bibitem [{\citenamefont {Escher}\ \emph {et~al.}(2011)\citenamefont {Escher},
  \citenamefont {{de Matos Filho}},\ and\ \citenamefont {Davidovich}}]{EMD11}%
  \BibitemOpen
  \bibfield  {author} {\bibinfo {author} {\bibfnamefont {B.~M.}\ \bibnamefont
  {Escher}}, \bibinfo {author} {\bibfnamefont {R.~L.}\ \bibnamefont {{de Matos
  Filho}}}, \ and\ \bibinfo {author} {\bibfnamefont {L.}~\bibnamefont
  {Davidovich}},\ }\href {\doibase 10.1038/nphys1958} {\bibfield  {journal}
  {\bibinfo  {journal} {Nat. Phys.}\ }\textbf {\bibinfo {volume} {7}},\
  \bibinfo {pages} {406} (\bibinfo {year} {2011})}\BibitemShut {NoStop}%
\bibitem [{\citenamefont {Demkowicz-Dobrzanski}\ \emph
  {et~al.}(2012)\citenamefont {Demkowicz-Dobrzanski}, \citenamefont
  {Ko{\l}odynski},\ and\ \citenamefont {Gu\c{t}\u{a}}}]{DKG12}%
  \BibitemOpen
  \bibfield  {author} {\bibinfo {author} {\bibfnamefont {R.}~\bibnamefont
  {Demkowicz-Dobrzanski}}, \bibinfo {author} {\bibfnamefont {J.}~\bibnamefont
  {Ko{\l}odynski}}, \ and\ \bibinfo {author} {\bibfnamefont {M.}~\bibnamefont
  {Gu\c{t}\u{a}}},\ }\href {\doibase 10.1038/ncomms2067} {\bibfield  {journal}
  {\bibinfo  {journal} {Nat. Commun.}\ }\textbf {\bibinfo {volume} {3}},\
  \bibinfo {pages} {1063} (\bibinfo {year} {2012})}\BibitemShut {NoStop}%
\bibitem [{\citenamefont {Berry}\ \emph {et~al.}(2001)\citenamefont {Berry},
  \citenamefont {Wiseman},\ and\ \citenamefont {Breslin}}]{BWB01}%
  \BibitemOpen
  \bibfield  {author} {\bibinfo {author} {\bibfnamefont {D.~W.}\ \bibnamefont
  {Berry}}, \bibinfo {author} {\bibfnamefont {H.~M.}\ \bibnamefont {Wiseman}},
  \ and\ \bibinfo {author} {\bibfnamefont {J.~K.}\ \bibnamefont {Breslin}},\
  }\href {\doibase 10.1103/PhysRevA.63.053804} {\bibfield  {journal} {\bibinfo
  {journal} {Phys. Rev. A}\ }\textbf {\bibinfo {volume} {63}},\ \bibinfo
  {pages} {053804} (\bibinfo {year} {2001})}\BibitemShut {NoStop}%
\bibitem [{\citenamefont {Bobroff}(1987)}]{Bob87}%
  \BibitemOpen
  \bibfield  {author} {\bibinfo {author} {\bibfnamefont {N.}~\bibnamefont
  {Bobroff}},\ }\href {\doibase 10.1364/AO.26.002676} {\bibfield  {journal}
  {\bibinfo  {journal} {Appl. Opt.}\ }\textbf {\bibinfo {volume} {26}},\
  \bibinfo {pages} {2676} (\bibinfo {year} {1987})}\BibitemShut {NoStop}%
\bibitem [{\citenamefont {Sinclair}\ \emph {et~al.}(2014)\citenamefont
  {Sinclair}, \citenamefont {Giorgetta}, \citenamefont {Swann}, \citenamefont
  {Baumann}, \citenamefont {Coddington},\ and\ \citenamefont
  {Newbury}}]{SGS+14}%
  \BibitemOpen
  \bibfield  {author} {\bibinfo {author} {\bibfnamefont {L.~C.}\ \bibnamefont
  {Sinclair}}, \bibinfo {author} {\bibfnamefont {F.~R.}\ \bibnamefont
  {Giorgetta}}, \bibinfo {author} {\bibfnamefont {W.~C.}\ \bibnamefont
  {Swann}}, \bibinfo {author} {\bibfnamefont {E.}~\bibnamefont {Baumann}},
  \bibinfo {author} {\bibfnamefont {I.}~\bibnamefont {Coddington}}, \ and\
  \bibinfo {author} {\bibfnamefont {N.~R.}\ \bibnamefont {Newbury}},\ }\href
  {\doibase 10.1103/PhysRevA.89.023805} {\bibfield  {journal} {\bibinfo
  {journal} {Phys. Rev. A}\ }\textbf {\bibinfo {volume} {89}},\ \bibinfo
  {pages} {023805} (\bibinfo {year} {2014})}\BibitemShut {NoStop}%
\bibitem [{\citenamefont {Severini}(2005{\natexlab{a}})}]{Sev05_ch12}%
  \BibitemOpen
  \bibfield  {author} {\bibinfo {author} {\bibfnamefont {T.~A.}\ \bibnamefont
  {Severini}},\ }\enquote {\bibinfo {title} {Central limit theorems},}\ in\
  \href {\doibase 10.1017/CBO9780511610547.013} {\emph {\bibinfo {booktitle}
  {Elements of Distribution Theory}}},\ \bibinfo {series and number} {Cambridge
  Series in Statistical and Probabilistic Mathematics}\ (\bibinfo  {publisher}
  {Cambridge University Press},\ \bibinfo {address} {Cambridge, UK},\ \bibinfo
  {year} {2005})\ Chap.~\bibinfo {chapter} {12}, pp.\ \bibinfo {pages}
  {365--399}\BibitemShut {NoStop}%
\bibitem [{\citenamefont {Breitenberger}(1963)}]{Bre63}%
  \BibitemOpen
  \bibfield  {author} {\bibinfo {author} {\bibfnamefont {E.}~\bibnamefont
  {Breitenberger}},\ }\href@noop {} {\bibfield  {journal} {\bibinfo  {journal}
  {Biometrika}\ }\textbf {\bibinfo {volume} {50}},\ \bibinfo {pages} {81}
  (\bibinfo {year} {1963})}\BibitemShut {NoStop}%
\bibitem [{\citenamefont {Newman}(1968)}]{New68}%
  \BibitemOpen
  \bibfield  {author} {\bibinfo {author} {\bibfnamefont {D.~S.}\ \bibnamefont
  {Newman}},\ }\href@noop {} {\bibfield  {journal} {\bibinfo  {journal} {Ann.
  Math. Stat.}\ }\textbf {\bibinfo {volume} {39}},\ \bibinfo {pages} {890}
  (\bibinfo {year} {1968})}\BibitemShut {NoStop}%
\bibitem [{\citenamefont {Azzalini}\ and\ \citenamefont
  {Capitanio}(2014)}]{AC14}%
  \BibitemOpen
  \bibfield  {author} {\bibinfo {author} {\bibfnamefont {A.}~\bibnamefont
  {Azzalini}}\ and\ \bibinfo {author} {\bibfnamefont {A.}~\bibnamefont
  {Capitanio}},\ }in\ \href@noop {} {\emph {\bibinfo {booktitle} {The
  Skew-Normal and Related Families}}},\ \bibinfo {series and number} {Institute
  of Mathematical Statistics Monograph},\ \bibinfo {editor} {edited by\
  \bibinfo {editor} {\bibfnamefont {D.~R.}\ \bibnamefont {Cox}}, \bibinfo
  {editor} {\bibfnamefont {A.}~\bibnamefont {Agresti}}, \bibinfo {editor}
  {\bibfnamefont {B.}~\bibnamefont {Hambly}}, \bibinfo {editor} {\bibfnamefont
  {S.}~\bibnamefont {Holmes}}, \ and\ \bibinfo {editor} {\bibfnamefont {X.-L.}\
  \bibnamefont {Meng}}}\ (\bibinfo  {publisher} {Cambridge University Press},\
  \bibinfo {address} {Cambridge, UK},\ \bibinfo {year} {2014})\ Chap.~\bibinfo
  {chapter} {2}, pp.\ \bibinfo {pages} {24 -- 56}\BibitemShut {NoStop}%
\bibitem [{\citenamefont {Limpert}\ \emph {et~al.}(2001)\citenamefont
  {Limpert}, \citenamefont {Stahel},\ and\ \citenamefont {Abbt}}]{LSA01}%
  \BibitemOpen
  \bibfield  {author} {\bibinfo {author} {\bibfnamefont {E.}~\bibnamefont
  {Limpert}}, \bibinfo {author} {\bibfnamefont {W.~A.}\ \bibnamefont {Stahel}},
  \ and\ \bibinfo {author} {\bibfnamefont {M.}~\bibnamefont {Abbt}},\
  }\href@noop {} {\bibfield  {journal} {\bibinfo  {journal} {BioScience}\
  }\textbf {\bibinfo {volume} {51}},\ \bibinfo {pages} {341} (\bibinfo {year}
  {2001})}\BibitemShut {NoStop}%
\bibitem [{\citenamefont {Pourahmadi}(2007)}]{Pou07}%
  \BibitemOpen
  \bibfield  {author} {\bibinfo {author} {\bibfnamefont {M.}~\bibnamefont
  {Pourahmadi}},\ }\href {\doibase 10.1080/03610920601126274} {\bibfield
  {journal} {\bibinfo  {journal} {Commun. Stat. Theory Methods}\ }\textbf
  {\bibinfo {volume} {36}},\ \bibinfo {pages} {1803} (\bibinfo {year}
  {2007})}\BibitemShut {NoStop}%
\bibitem [{\citenamefont {Jouini}(2011)}]{Jou11}%
  \BibitemOpen
  \bibfield  {author} {\bibinfo {author} {\bibfnamefont {W.}~\bibnamefont
  {Jouini}},\ }\href {\doibase 10.1109/LSP.2011.2155649} {\bibfield  {journal}
  {\bibinfo  {journal} {{IEEE} Signal Process. Lett.}\ }\textbf {\bibinfo
  {volume} {18}},\ \bibinfo {pages} {423} (\bibinfo {year} {2011})}\BibitemShut
  {NoStop}%
\bibitem [{\citenamefont {Rezaie}\ and\ \citenamefont {Eidsvik}(2014)}]{RE14}%
  \BibitemOpen
  \bibfield  {author} {\bibinfo {author} {\bibfnamefont {J.}~\bibnamefont
  {Rezaie}}\ and\ \bibinfo {author} {\bibfnamefont {J.}~\bibnamefont
  {Eidsvik}},\ }\href {\doibase https://doi.org/10.1016/j.csda.2014.01.014}
  {\bibfield  {journal} {\bibinfo  {journal} {Comput. Stat. Data Anal}\
  }\textbf {\bibinfo {volume} {75}},\ \bibinfo {pages} {1 } (\bibinfo {year}
  {2014})}\BibitemShut {NoStop}%
\bibitem [{\citenamefont {Vrajitoru}\ and\ \citenamefont
  {Knight}(2014{\natexlab{b}})}]{VK14_ch5}%
  \BibitemOpen
  \bibfield  {author} {\bibinfo {author} {\bibfnamefont {D.}~\bibnamefont
  {Vrajitoru}}\ and\ \bibinfo {author} {\bibfnamefont {W.}~\bibnamefont
  {Knight}},\ }\enquote {\bibinfo {title} {Deterministic analysis of
  algorithms},}\ in\ \href {\doibase 10.1007/978-3-319-09888-3_5} {\emph
  {\bibinfo {booktitle} {Practical Analysis of Algorithms}}}\ (\bibinfo
  {publisher} {Springer},\ \bibinfo {address} {Cham, Switzerland},\ \bibinfo
  {year} {2014})\ Chap.~\bibinfo {chapter} {5}, pp.\ \bibinfo {pages}
  {169--293}\BibitemShut {NoStop}%
\bibitem [{\citenamefont {Palittpongarnpim}\ \emph {et~al.}(2016)\citenamefont
  {Palittpongarnpim}, \citenamefont {Wittek},\ and\ \citenamefont
  {Sanders}}]{PWS16b}%
  \BibitemOpen
  \bibfield  {author} {\bibinfo {author} {\bibfnamefont {P.}~\bibnamefont
  {Palittpongarnpim}}, \bibinfo {author} {\bibfnamefont {P.}~\bibnamefont
  {Wittek}}, \ and\ \bibinfo {author} {\bibfnamefont {B.~C.}\ \bibnamefont
  {Sanders}},\ }in\ \href {\doibase 10.1117/12.2237355} {\emph {\bibinfo
  {booktitle} {Proc. {SPIE} Quantum Communications and Quantum Imaging XIV}}},\
  Vol.\ \bibinfo {volume} {9980}\ (\bibinfo  {publisher} {{SPIE}},\ \bibinfo
  {address} {Bellingham, WA},\ \bibinfo {year} {2016})\ pp.\ \bibinfo {pages}
  {99800H--{99800H--11}}\BibitemShut {NoStop}%
\bibitem [{\citenamefont {Helstrom}(1969)}]{Hel69}%
  \BibitemOpen
  \bibfield  {author} {\bibinfo {author} {\bibfnamefont {C.~W.}\ \bibnamefont
  {Helstrom}},\ }\href {\doibase 10.1007/BF01007479} {\bibfield  {journal}
  {\bibinfo  {journal} {J. Stat. Phys.}\ }\textbf {\bibinfo {volume} {1}},\
  \bibinfo {pages} {231} (\bibinfo {year} {1969})}\BibitemShut {NoStop}%
\bibitem [{\citenamefont {Humphreys}\ \emph {et~al.}(2013)\citenamefont
  {Humphreys}, \citenamefont {Barbieri}, \citenamefont {Datta},\ and\
  \citenamefont {Walmsley}}]{HBDW13}%
  \BibitemOpen
  \bibfield  {author} {\bibinfo {author} {\bibfnamefont {P.~C.}\ \bibnamefont
  {Humphreys}}, \bibinfo {author} {\bibfnamefont {M.}~\bibnamefont {Barbieri}},
  \bibinfo {author} {\bibfnamefont {A.}~\bibnamefont {Datta}}, \ and\ \bibinfo
  {author} {\bibfnamefont {I.~A.}\ \bibnamefont {Walmsley}},\ }\href {\doibase
  10.1103/PhysRevLett.111.070403} {\bibfield  {journal} {\bibinfo  {journal}
  {Phys. Rev. Lett.}\ }\textbf {\bibinfo {volume} {111}},\ \bibinfo {pages}
  {070403} (\bibinfo {year} {2013})}\BibitemShut {NoStop}%
\bibitem [{\citenamefont {Yue}\ \emph {et~al.}(2014)\citenamefont {Yue},
  \citenamefont {Zhang},\ and\ \citenamefont {Fan}}]{YZF14}%
  \BibitemOpen
  \bibfield  {author} {\bibinfo {author} {\bibfnamefont {J.-D.}\ \bibnamefont
  {Yue}}, \bibinfo {author} {\bibfnamefont {Y.-R.}\ \bibnamefont {Zhang}}, \
  and\ \bibinfo {author} {\bibfnamefont {H.}~\bibnamefont {Fan}},\ }\href
  {\doibase 10.1038/srep05933} {\bibfield  {journal} {\bibinfo  {journal} {Sci.
  Rep.}\ }\textbf {\bibinfo {volume} {4}},\ \bibinfo {pages} {5933} (\bibinfo
  {year} {2014})}\BibitemShut {NoStop}%
\bibitem [{\citenamefont {Higgins}\ \emph {et~al.}(2007)\citenamefont
  {Higgins}, \citenamefont {Berry}, \citenamefont {Bartlett}, \citenamefont
  {Wiseman},\ and\ \citenamefont {Pryde}}]{HBB+07}%
  \BibitemOpen
  \bibfield  {author} {\bibinfo {author} {\bibfnamefont {B.~L.}\ \bibnamefont
  {Higgins}}, \bibinfo {author} {\bibfnamefont {D.~W.}\ \bibnamefont {Berry}},
  \bibinfo {author} {\bibfnamefont {S.~D.}\ \bibnamefont {Bartlett}}, \bibinfo
  {author} {\bibfnamefont {H.~M.}\ \bibnamefont {Wiseman}}, \ and\ \bibinfo
  {author} {\bibfnamefont {G.~J.}\ \bibnamefont {Pryde}},\ }\href {\doibase
  10.1038/nature06257} {\bibfield  {journal} {\bibinfo  {journal} {Nature}\
  }\textbf {\bibinfo {volume} {450}},\ \bibinfo {pages} {393} (\bibinfo {year}
  {2007})}\BibitemShut {NoStop}%
\bibitem [{\citenamefont {Demkowicz-Dobrza\'{n}ski}\ and\ \citenamefont
  {Maccone}(2014)}]{DM14}%
  \BibitemOpen
  \bibfield  {author} {\bibinfo {author} {\bibfnamefont {R.}~\bibnamefont
  {Demkowicz-Dobrza\'{n}ski}}\ and\ \bibinfo {author} {\bibfnamefont
  {L.}~\bibnamefont {Maccone}},\ }\href {\doibase
  10.1103/PhysRevLett.113.250801} {\bibfield  {journal} {\bibinfo  {journal}
  {Phys. Rev. Lett.}\ }\textbf {\bibinfo {volume} {113}},\ \bibinfo {pages}
  {250801} (\bibinfo {year} {2014})}\BibitemShut {NoStop}%
\bibitem [{\citenamefont {Hotta}\ \emph {et~al.}(2005)\citenamefont {Hotta},
  \citenamefont {Karasawa},\ and\ \citenamefont {Ozawa}}]{HKO05}%
  \BibitemOpen
  \bibfield  {author} {\bibinfo {author} {\bibfnamefont {M.}~\bibnamefont
  {Hotta}}, \bibinfo {author} {\bibfnamefont {T.}~\bibnamefont {Karasawa}}, \
  and\ \bibinfo {author} {\bibfnamefont {M.}~\bibnamefont {Ozawa}},\ }\href
  {\doibase 10.1103/PhysRevA.72.052334} {\bibfield  {journal} {\bibinfo
  {journal} {Phys. Rev. A}\ }\textbf {\bibinfo {volume} {72}},\ \bibinfo
  {pages} {052334} (\bibinfo {year} {2005})}\BibitemShut {NoStop}%
\bibitem [{\citenamefont {Sbroscia}\ \emph {et~al.}(2018)\citenamefont
  {Sbroscia}, \citenamefont {Gianani}, \citenamefont {Mancino}, \citenamefont
  {Roccia}, \citenamefont {Huang}, \citenamefont {Maccone}, \citenamefont
  {Macchiavello},\ and\ \citenamefont {Barbieri}}]{SGM+18}%
  \BibitemOpen
  \bibfield  {author} {\bibinfo {author} {\bibfnamefont {M.}~\bibnamefont
  {Sbroscia}}, \bibinfo {author} {\bibfnamefont {I.}~\bibnamefont {Gianani}},
  \bibinfo {author} {\bibfnamefont {L.}~\bibnamefont {Mancino}}, \bibinfo
  {author} {\bibfnamefont {E.}~\bibnamefont {Roccia}}, \bibinfo {author}
  {\bibfnamefont {Z.}~\bibnamefont {Huang}}, \bibinfo {author} {\bibfnamefont
  {L.}~\bibnamefont {Maccone}}, \bibinfo {author} {\bibfnamefont
  {C.}~\bibnamefont {Macchiavello}}, \ and\ \bibinfo {author} {\bibfnamefont
  {M.}~\bibnamefont {Barbieri}},\ }\href {\doibase 10.1103/PhysRevA.97.032305}
  {\bibfield  {journal} {\bibinfo  {journal} {Phys. Rev. A}\ }\textbf {\bibinfo
  {volume} {97}},\ \bibinfo {pages} {032305} (\bibinfo {year}
  {2018})}\BibitemShut {NoStop}%
\bibitem [{\citenamefont {Landsman}(2017)}]{Lan17}%
  \BibitemOpen
  \bibfield  {author} {\bibinfo {author} {\bibfnamefont {K.}~\bibnamefont
  {Landsman}},\ }\enquote {\bibinfo {title} {Quantum physics on a general
  hilbert space},}\ in\ \href {\doibase 10.1007/978-3-319-51777-3_4} {\emph
  {\bibinfo {booktitle} {Foundations of Quantum Theory: From Classical Concepts
  to Operator Algebras}}}\ (\bibinfo  {publisher} {Springer International},\
  \bibinfo {address} {Cham, Switzerland},\ \bibinfo {year} {2017})\
  Chap.~\bibinfo {chapter} {4}, pp.\ \bibinfo {pages} {103--123}\BibitemShut
  {NoStop}%
\bibitem [{\citenamefont {Holevo}(2012)}]{Hol12}%
  \BibitemOpen
  \bibfield  {author} {\bibinfo {author} {\bibfnamefont {A.~S.}\ \bibnamefont
  {Holevo}},\ }in\ \href {\doibase 10.1515/9783110273403} {\emph {\bibinfo
  {booktitle} {Quantum Systems, Channels, Information: A Mathematical
  Introduction}}},\ \bibinfo {series} {De Gruyter Studies in Mathematical
  Physics}, Vol.~\bibinfo {volume} {16}\ (\bibinfo  {publisher} {Walter de
  Gruyter},\ \bibinfo {address} {Berlin, Germany},\ \bibinfo {year} {2012})\
  Chap.\ \bibinfo {chapter} {Quantum evolutions and channels}, pp.\ \bibinfo
  {pages} {103--131}\BibitemShut {NoStop}%
\bibitem [{\citenamefont {Yurke}\ \emph {et~al.}(1986)\citenamefont {Yurke},
  \citenamefont {McCall},\ and\ \citenamefont {Klauder}}]{YMK86}%
  \BibitemOpen
  \bibfield  {author} {\bibinfo {author} {\bibfnamefont {B.}~\bibnamefont
  {Yurke}}, \bibinfo {author} {\bibfnamefont {S.~L.}\ \bibnamefont {McCall}}, \
  and\ \bibinfo {author} {\bibfnamefont {J.~R.}\ \bibnamefont {Klauder}},\
  }\href {\doibase 10.1103/PhysRevA.33.4033} {\bibfield  {journal} {\bibinfo
  {journal} {Phys. Rev. A}\ }\textbf {\bibinfo {volume} {33}},\ \bibinfo
  {pages} {4033} (\bibinfo {year} {1986})}\BibitemShut {NoStop}%
\bibitem [{\citenamefont {Hayashi}(2006)}]{Hay2006}%
  \BibitemOpen
  \bibfield  {author} {\bibinfo {author} {\bibfnamefont {M.}~\bibnamefont
  {Hayashi}},\ }\enquote {\bibinfo {title} {Mathematical formulation of quantum
  systems},}\ in\ \href {\doibase 10.1007/3-540-30266-2_2} {\emph {\bibinfo
  {booktitle} {Quantum Information: An Introduction}}}\ (\bibinfo  {publisher}
  {Springer},\ \bibinfo {address} {Berlin, Germany},\ \bibinfo {year} {2006})\
  pp.\ \bibinfo {pages} {9--25}\BibitemShut {NoStop}%
\bibitem [{\citenamefont {Wiseman}\ and\ \citenamefont {Milburn}(2009)}]{WM09}%
  \BibitemOpen
  \bibfield  {author} {\bibinfo {author} {\bibfnamefont {H.~M.}\ \bibnamefont
  {Wiseman}}\ and\ \bibinfo {author} {\bibfnamefont {G.~J.}\ \bibnamefont
  {Milburn}},\ }\href@noop {} {\emph {\bibinfo {title} {Quantum Measurement and
  Control}}}\ (\bibinfo  {publisher} {Cambridge University Press},\ \bibinfo
  {address} {Cambridge, MA},\ \bibinfo {year} {2009})\BibitemShut {NoStop}%
\bibitem [{\citenamefont {James}\ and\ \citenamefont {Nurdin}(2015)}]{JN15}%
  \BibitemOpen
  \bibfield  {author} {\bibinfo {author} {\bibfnamefont {M.~R.}\ \bibnamefont
  {James}}\ and\ \bibinfo {author} {\bibfnamefont {H.~I.}\ \bibnamefont
  {Nurdin}},\ }in\ \href {\doibase 10.1109/CCA.2015.7320607} {\emph {\bibinfo
  {booktitle} {2015 IEEE Conference on Control Applications (CCA)}}},\ \bibinfo
  {organization} {{IEEE}}\ (\bibinfo  {publisher} {{IEEE CCA}},\ \bibinfo
  {address} {Sydney, NSW},\ \bibinfo {year} {2015})\ pp.\ \bibinfo {pages}
  {1--12}\BibitemShut {NoStop}%
\bibitem [{\citenamefont {Schreppler}\ \emph {et~al.}(2014)\citenamefont
  {Schreppler}, \citenamefont {Spethmann}, \citenamefont {Brahms},
  \citenamefont {Botter}, \citenamefont {Barrios},\ and\ \citenamefont
  {Stamper-Kurn}}]{Sch14}%
  \BibitemOpen
  \bibfield  {author} {\bibinfo {author} {\bibfnamefont {S.}~\bibnamefont
  {Schreppler}}, \bibinfo {author} {\bibfnamefont {N.}~\bibnamefont
  {Spethmann}}, \bibinfo {author} {\bibfnamefont {N.}~\bibnamefont {Brahms}},
  \bibinfo {author} {\bibfnamefont {T.}~\bibnamefont {Botter}}, \bibinfo
  {author} {\bibfnamefont {M.}~\bibnamefont {Barrios}}, \ and\ \bibinfo
  {author} {\bibfnamefont {D.~M.}\ \bibnamefont {Stamper-Kurn}},\ }\href
  {\doibase 10.1126/science.1249850} {\bibfield  {journal} {\bibinfo  {journal}
  {Science}\ }\textbf {\bibinfo {volume} {344}},\ \bibinfo {pages} {1486}
  (\bibinfo {year} {2014})}\BibitemShut {NoStop}%
\bibitem [{\citenamefont {Ko\l{}ody\'{n}ski{}}\ and\ \citenamefont
  {Demkowicz-Dobrza\'{n}ski}(2013)}]{KD13}%
  \BibitemOpen
  \bibfield  {author} {\bibinfo {author} {\bibfnamefont {J.}~\bibnamefont
  {Ko\l{}ody\'{n}ski{}}}\ and\ \bibinfo {author} {\bibfnamefont
  {R.}~\bibnamefont {Demkowicz-Dobrza\'{n}ski}},\ }\href {\doibase
  10.1088/1367-2630/15/7/073043} {\bibfield  {journal} {\bibinfo  {journal}
  {New J. Phys.}\ }\textbf {\bibinfo {volume} {15}},\ \bibinfo {pages} {073043}
  (\bibinfo {year} {2013})}\BibitemShut {NoStop}%
\bibitem [{\citenamefont {Maccone}(2013)}]{Mac13}%
  \BibitemOpen
  \bibfield  {author} {\bibinfo {author} {\bibfnamefont {L.}~\bibnamefont
  {Maccone}},\ }\href {\doibase 10.1103/PhysRevA.88.042109} {\bibfield
  {journal} {\bibinfo  {journal} {Phys. Rev. A}\ }\textbf {\bibinfo {volume}
  {88}},\ \bibinfo {pages} {042109} (\bibinfo {year} {2013})}\BibitemShut
  {NoStop}%
\bibitem [{\citenamefont {Kay}(1993)}]{Kay93}%
  \BibitemOpen
  \bibfield  {author} {\bibinfo {author} {\bibfnamefont {S.~M.}\ \bibnamefont
  {Kay}},\ }\enquote {\bibinfo {title} {{C}ramer-{R}ao lower bound},}\ in\
  \href@noop {} {\emph {\bibinfo {booktitle} {Fundamentals of Statistical
  Signal Processing: Estimation Theory}}},\ \bibinfo {editor} {edited by\
  \bibinfo {editor} {\bibfnamefont {A.~V.}\ \bibnamefont {Oppenheimer}}}\
  (\bibinfo  {publisher} {Prentice-Hall},\ \bibinfo {address} {Upper Saddle
  River, NJ},\ \bibinfo {year} {1993})\ Chap.~\bibinfo {chapter}
  {3}\BibitemShut {NoStop}%
\bibitem [{\citenamefont {Braunstein}(2005)}]{Brau05}%
  \BibitemOpen
  \bibfield  {author} {\bibinfo {author} {\bibfnamefont {S.~L.}\ \bibnamefont
  {Braunstein}},\ }\href {\doibase 10.1103/PhysRevA.71.055801} {\bibfield
  {journal} {\bibinfo  {journal} {Phys. Rev. A}\ }\textbf {\bibinfo {volume}
  {71}},\ \bibinfo {pages} {055801} (\bibinfo {year} {2005})}\BibitemShut
  {NoStop}%
\bibitem [{\citenamefont {Wootters}(1998)}]{Woo98}%
  \BibitemOpen
  \bibfield  {author} {\bibinfo {author} {\bibfnamefont {W.~K.}\ \bibnamefont
  {Wootters}},\ }\href {\doibase 10.1098/rsta.1998.0244} {\bibfield  {journal}
  {\bibinfo  {journal} {Philos. Trans. Royal Soc. A: Math. Phys. Eng. Sci.}\
  }\textbf {\bibinfo {volume} {356}},\ \bibinfo {pages} {1717} (\bibinfo {year}
  {1998})}\BibitemShut {NoStop}%
\bibitem [{\citenamefont {Jacobs}(2014)}]{Jacobs2014_ch6}%
  \BibitemOpen
  \bibfield  {author} {\bibinfo {author} {\bibfnamefont {K.}~\bibnamefont
  {Jacobs}},\ }\enquote {\bibinfo {title} {Metrology},}\ in\ \href@noop {}
  {\emph {\bibinfo {booktitle} {Qauntum Measurement Theory and its
  Applications}}}\ (\bibinfo  {publisher} {Cambridge University Press},\
  \bibinfo {address} {Cambridge, UK},\ \bibinfo {year} {2014})\ Chap.~\bibinfo
  {chapter} {6}, pp.\ \bibinfo {pages} {303--322}\BibitemShut {NoStop}%
\bibitem [{\citenamefont {Zwierz}\ \emph {et~al.}(2012)\citenamefont {Zwierz},
  \citenamefont {{P\'erez-Delgado}},\ and\ \citenamefont {Kok}}]{ZPK12}%
  \BibitemOpen
  \bibfield  {author} {\bibinfo {author} {\bibfnamefont {M.}~\bibnamefont
  {Zwierz}}, \bibinfo {author} {\bibfnamefont {C.~A.}\ \bibnamefont
  {{P\'erez-Delgado}}}, \ and\ \bibinfo {author} {\bibfnamefont
  {P.}~\bibnamefont {Kok}},\ }\href {\doibase 10.1103/PhysRevA.85.042112}
  {\bibfield  {journal} {\bibinfo  {journal} {Phys. Rev. A}\ }\textbf {\bibinfo
  {volume} {85}},\ \bibinfo {pages} {042112} (\bibinfo {year}
  {2012})}\BibitemShut {NoStop}%
\bibitem [{\citenamefont {Bondurant}\ and\ \citenamefont
  {Shapiro}(1984)}]{BS84}%
  \BibitemOpen
  \bibfield  {author} {\bibinfo {author} {\bibfnamefont {R.~S.}\ \bibnamefont
  {Bondurant}}\ and\ \bibinfo {author} {\bibfnamefont {J.~H.}\ \bibnamefont
  {Shapiro}},\ }\href {\doibase 10.1103/PhysRevD.30.2548} {\bibfield  {journal}
  {\bibinfo  {journal} {Phys. Rev. D}\ }\textbf {\bibinfo {volume} {30}},\
  \bibinfo {pages} {2548} (\bibinfo {year} {1984})}\BibitemShut {NoStop}%
\bibitem [{\citenamefont {Wiseman}\ \emph {et~al.}(2009)\citenamefont
  {Wiseman}, \citenamefont {Berry}, \citenamefont {Bartlett}, \citenamefont
  {Higgins},\ and\ \citenamefont {Pryde}}]{WBB+09}%
  \BibitemOpen
  \bibfield  {author} {\bibinfo {author} {\bibfnamefont {H.~M.}\ \bibnamefont
  {Wiseman}}, \bibinfo {author} {\bibfnamefont {D.~W.}\ \bibnamefont {Berry}},
  \bibinfo {author} {\bibfnamefont {S.~D.}\ \bibnamefont {Bartlett}}, \bibinfo
  {author} {\bibfnamefont {B.~L.}\ \bibnamefont {Higgins}}, \ and\ \bibinfo
  {author} {\bibfnamefont {G.~J.}\ \bibnamefont {Pryde}},\ }\href {\doibase
  10.1109/JSTQE.2009.2020810} {\bibfield  {journal} {\bibinfo  {journal} {IEEE
  J. Sel. Top. Quantum Electron.}\ }\textbf {\bibinfo {volume} {15}},\ \bibinfo
  {pages} {1661} (\bibinfo {year} {2009})}\BibitemShut {NoStop}%
\bibitem [{\citenamefont {Hentschel}\ and\ \citenamefont
  {Sanders}(2011{\natexlab{a}})}]{HS11b}%
  \BibitemOpen
  \bibfield  {author} {\bibinfo {author} {\bibfnamefont {A.}~\bibnamefont
  {Hentschel}}\ and\ \bibinfo {author} {\bibfnamefont {B.~C.}\ \bibnamefont
  {Sanders}},\ }\href {\doibase 10.1103/PhysRevLett.107.233601} {\bibfield
  {journal} {\bibinfo  {journal} {Phys. Rev. Lett.}\ }\textbf {\bibinfo
  {volume} {107}},\ \bibinfo {pages} {233601} (\bibinfo {year}
  {2011}{\natexlab{a}})}\BibitemShut {NoStop}%
\bibitem [{\citenamefont {Armstrong}(1966)}]{Arm66}%
  \BibitemOpen
  \bibfield  {author} {\bibinfo {author} {\bibfnamefont {J.~A.}\ \bibnamefont
  {Armstrong}},\ }\href {\doibase 10.1364/JOSA.56.001024} {\bibfield  {journal}
  {\bibinfo  {journal} {J. Opt. Soc. Am.}\ }\textbf {\bibinfo {volume} {56}},\
  \bibinfo {pages} {1024} (\bibinfo {year} {1966})}\BibitemShut {NoStop}%
\bibitem [{\citenamefont {Hariharan}\ and\ \citenamefont
  {Sanders}(1996)}]{HS96}%
  \BibitemOpen
  \bibfield  {author} {\bibinfo {author} {\bibfnamefont {P.}~\bibnamefont
  {Hariharan}}\ and\ \bibinfo {author} {\bibfnamefont {B.~C.}\ \bibnamefont
  {Sanders}},\ }in\ \href@noop {} {\emph {\bibinfo {booktitle} {Progress in
  Optics}}},\ Vol.\ \bibinfo {volume} {XXXVI},\ \bibinfo {editor} {edited by\
  \bibinfo {editor} {\bibfnamefont {E.}~\bibnamefont {Wolf}}}\ (\bibinfo
  {publisher} {Elsevier},\ \bibinfo {address} {Amsterdam, Netherlands},\
  \bibinfo {year} {1996})\ Chap.~\bibinfo {chapter} {2}, pp.\ \bibinfo {pages}
  {49--128}\BibitemShut {NoStop}%
\bibitem [{\citenamefont {Hentschel}\ and\ \citenamefont
  {Sanders}(2011{\natexlab{b}})}]{HS11a}%
  \BibitemOpen
  \bibfield  {author} {\bibinfo {author} {\bibfnamefont {A.}~\bibnamefont
  {Hentschel}}\ and\ \bibinfo {author} {\bibfnamefont {B.~C.}\ \bibnamefont
  {Sanders}},\ }\href {\doibase 10.1088/1751-8113/44/11/115301} {\bibfield
  {journal} {\bibinfo  {journal} {J. Phys. A: Math. Theor.}\ }\textbf {\bibinfo
  {volume} {44}},\ \bibinfo {pages} {115301} (\bibinfo {year}
  {2011}{\natexlab{b}})}\BibitemShut {NoStop}%
\bibitem [{\citenamefont {Summy}\ and\ \citenamefont {Pegg}(1990)}]{SP90}%
  \BibitemOpen
  \bibfield  {author} {\bibinfo {author} {\bibfnamefont {G.}~\bibnamefont
  {Summy}}\ and\ \bibinfo {author} {\bibfnamefont {D.}~\bibnamefont {Pegg}},\
  }\href {\doibase https://doi.org/10.1016/0030-4018(90)90464-5} {\bibfield
  {journal} {\bibinfo  {journal} {Opt. Commun.}\ }\textbf {\bibinfo {volume}
  {77}},\ \bibinfo {pages} {75} (\bibinfo {year} {1990})}\BibitemShut {NoStop}%
\bibitem [{\citenamefont {Mishchenko}(2014)}]{Mis14}%
  \BibitemOpen
  \bibfield  {author} {\bibinfo {author} {\bibfnamefont {M.~I.}\ \bibnamefont
  {Mishchenko}},\ }\enquote {\bibinfo {title} {Wigner d-functions},}\ in\ \href
  {\doibase 10.1017/CBO9781139019064.029} {\emph {\bibinfo {booktitle}
  {Electromagnetic Scattering by Particles and Particle Groups: An
  Introduction}}}\ (\bibinfo  {publisher} {Cambridge University Press},\
  \bibinfo {year} {2014})\ Chap.\ \bibinfo {chapter} {Appendix F}, pp.\
  \bibinfo {pages} {385--389}\BibitemShut {NoStop}%
\bibitem [{\citenamefont {Sanders}\ and\ \citenamefont {Milburn}(1995)}]{SM95}%
  \BibitemOpen
  \bibfield  {author} {\bibinfo {author} {\bibfnamefont {B.~C.}\ \bibnamefont
  {Sanders}}\ and\ \bibinfo {author} {\bibfnamefont {G.~J.}\ \bibnamefont
  {Milburn}},\ }\href {\doibase 10.1103/PhysRevLett.75.2944} {\bibfield
  {journal} {\bibinfo  {journal} {Phys. Rev. Lett.}\ }\textbf {\bibinfo
  {volume} {75}},\ \bibinfo {pages} {2944} (\bibinfo {year}
  {1995})}\BibitemShut {NoStop}%
\bibitem [{\citenamefont {Riley}\ \emph
  {et~al.}(2006{\natexlab{a}})\citenamefont {Riley}, \citenamefont {Hobson},\
  and\ \citenamefont {J.}}]{RHB06_ch30}%
  \BibitemOpen
  \bibfield  {author} {\bibinfo {author} {\bibfnamefont {K.~F.}\ \bibnamefont
  {Riley}}, \bibinfo {author} {\bibfnamefont {M.~P.}\ \bibnamefont {Hobson}}, \
  and\ \bibinfo {author} {\bibfnamefont {B.~S.}\ \bibnamefont {J.}},\ }\enquote
  {\bibinfo {title} {Probability},}\ \ (\bibinfo  {publisher} {Cambridge
  University Press},\ \bibinfo {address} {Cambridge, UK},\ \bibinfo {year}
  {2006})\ Chap.~\bibinfo {chapter} {30}, pp.\ \bibinfo {pages} {1119--1220},\
  \bibinfo {edition} {3rd}\ ed.\BibitemShut {Stop}%
\bibitem [{\citenamefont {Leigh}(2004)}]{Lei04_ch6}%
  \BibitemOpen
  \bibfield  {author} {\bibinfo {author} {\bibfnamefont {J.~R.}\ \bibnamefont
  {Leigh}},\ }\enquote {\bibinfo {title} {Mathematical modelling},}\ in\
  \href@noop {} {\emph {\bibinfo {booktitle} {Control Theory (2nd Edition)}}}\
  (\bibinfo  {publisher} {Institution of Engineering and Technology},\ \bibinfo
  {address} {London, UK},\ \bibinfo {year} {2004})\ Chap.~\bibinfo {chapter}
  {6}, pp.\ \bibinfo {pages} {61--81},\ \bibinfo {edition} {2nd}\
  ed.\BibitemShut {Stop}%
\bibitem [{\citenamefont {Wang}\ \emph {et~al.}(2016)\citenamefont {Wang},
  \citenamefont {Simkoff}, \citenamefont {Baldea}, \citenamefont {Chiang},
  \citenamefont {Castillo}, \citenamefont {Bindlish},\ and\ \citenamefont
  {Stanley}}]{WSB+16}%
  \BibitemOpen
  \bibfield  {author} {\bibinfo {author} {\bibfnamefont {S.}~\bibnamefont
  {Wang}}, \bibinfo {author} {\bibfnamefont {J.~M.}\ \bibnamefont {Simkoff}},
  \bibinfo {author} {\bibfnamefont {M.}~\bibnamefont {Baldea}}, \bibinfo
  {author} {\bibfnamefont {L.~H.}\ \bibnamefont {Chiang}}, \bibinfo {author}
  {\bibfnamefont {I.}~\bibnamefont {Castillo}}, \bibinfo {author}
  {\bibfnamefont {R.}~\bibnamefont {Bindlish}}, \ and\ \bibinfo {author}
  {\bibfnamefont {D.~B.}\ \bibnamefont {Stanley}},\ }in\ \href {\doibase
  10.1016/j.ifacol.2016.07.211} {\emph {\bibinfo {booktitle} {11th IFAC
  Symposium on Dynamics and Control of Process Systems Including Biosystems
  2016}}},\ Vol.~\bibinfo {volume} {49}\ (\bibinfo  {publisher} {Elsevier},\
  \bibinfo {address} {Amsterdam, Netherlands},\ \bibinfo {year} {2016})\ pp.\
  \bibinfo {pages} {25--30}\BibitemShut {NoStop}%
\bibitem [{\citenamefont {Hou}\ and\ \citenamefont {Wang}(2013)}]{HW13}%
  \BibitemOpen
  \bibfield  {author} {\bibinfo {author} {\bibfnamefont {Z.-S.}\ \bibnamefont
  {Hou}}\ and\ \bibinfo {author} {\bibfnamefont {Z.}~\bibnamefont {Wang}},\
  }\href {\doibase https://doi.org/10.1016/j.ins.2012.07.014} {\bibfield
  {journal} {\bibinfo  {journal} {Information Sciences}\ }\textbf {\bibinfo
  {volume} {235}},\ \bibinfo {pages} {3} (\bibinfo {year} {2013})}\BibitemShut
  {NoStop}%
\bibitem [{\citenamefont {Sutton}\ and\ \citenamefont
  {Barto}(2017)}]{SB2017_ch1}%
  \BibitemOpen
  \bibfield  {author} {\bibinfo {author} {\bibfnamefont {R.~S.}\ \bibnamefont
  {Sutton}}\ and\ \bibinfo {author} {\bibfnamefont {A.~G.}\ \bibnamefont
  {Barto}},\ }\enquote {\bibinfo {title} {Introduction},}\ in\ \href@noop {}
  {\emph {\bibinfo {booktitle} {Reinforcement Learning: An Introduction}}},\
  \bibinfo {series and number} {Adaptive Computation and Machine Learning}\
  (\bibinfo  {publisher} {MIT},\ \bibinfo {address} {Massachusetts},\ \bibinfo
  {year} {2017})\ Chap.~\bibinfo {chapter} {1}, pp.\ \bibinfo {pages} {3--24},\
  \bibinfo {edition} {2nd}\ ed.\BibitemShut {Stop}%
\bibitem [{\citenamefont {Sutton}\ \emph {et~al.}(1992)\citenamefont {Sutton},
  \citenamefont {Barto},\ and\ \citenamefont {Williams}}]{SBW92}%
  \BibitemOpen
  \bibfield  {author} {\bibinfo {author} {\bibfnamefont {R.~S.}\ \bibnamefont
  {Sutton}}, \bibinfo {author} {\bibfnamefont {A.~G.}\ \bibnamefont {Barto}}, \
  and\ \bibinfo {author} {\bibfnamefont {R.~J.}\ \bibnamefont {Williams}},\
  }\href {\doibase 10.1109/37.126844} {\bibfield  {journal} {\bibinfo
  {journal} {{IEEE} Control Systems}\ }\textbf {\bibinfo {volume} {12}},\
  \bibinfo {pages} {19} (\bibinfo {year} {1992})}\BibitemShut {NoStop}%
\bibitem [{\citenamefont {Degris}\ \emph {et~al.}(2012)\citenamefont {Degris},
  \citenamefont {Pilarski},\ and\ \citenamefont {Sutton}}]{DPS12}%
  \BibitemOpen
  \bibfield  {author} {\bibinfo {author} {\bibfnamefont {T.}~\bibnamefont
  {Degris}}, \bibinfo {author} {\bibfnamefont {P.~M.}\ \bibnamefont
  {Pilarski}}, \ and\ \bibinfo {author} {\bibfnamefont {R.~S.}\ \bibnamefont
  {Sutton}},\ }in\ \href {\doibase 10.1109/ACC.2012.6315022} {\emph {\bibinfo
  {booktitle} {2012 American Control Conference (ACC)}}},\ \bibinfo
  {organization} {{IEEE}}\ (\bibinfo  {publisher} {{IEEE}},\ \bibinfo {address}
  {Montreal, Canada},\ \bibinfo {year} {2012})\ pp.\ \bibinfo {pages}
  {2177--2182}\BibitemShut {NoStop}%
\bibitem [{\citenamefont {Severini}(2005{\natexlab{b}})}]{Sev05_ch4}%
  \BibitemOpen
  \bibfield  {author} {\bibinfo {author} {\bibfnamefont {T.~A.}\ \bibnamefont
  {Severini}},\ }\enquote {\bibinfo {title} {Moments and cumulants},}\ in\
  \href {\doibase 10.1017/CBO9780511610547.005} {\emph {\bibinfo {booktitle}
  {Elements of Distribution Theory}}},\ \bibinfo {series and number} {Cambridge
  Series in Statistical and Probabilistic Mathematics}\ (\bibinfo  {publisher}
  {Cambridge University Press},\ \bibinfo {address} {Cambridge, UK},\ \bibinfo
  {year} {2005})\ Chap.~\bibinfo {chapter} {4}, pp.\ \bibinfo {pages}
  {94--131}\BibitemShut {NoStop}%
\bibitem [{\citenamefont {Kirkup}(2012)}]{Kir12_ch3}%
  \BibitemOpen
  \bibfield  {author} {\bibinfo {author} {\bibfnamefont {L.}~\bibnamefont
  {Kirkup}},\ }\enquote {\bibinfo {title} {Data distributions i},}\ in\ \href
  {\doibase 10.1017/CBO9781139005258.005} {\emph {\bibinfo {booktitle} {Data
  Analysis for Physical Scientists: Featuring Excel{\textregistered}}}}\
  (\bibinfo  {publisher} {Cambridge University Press},\ \bibinfo {address}
  {Cambridge, UK},\ \bibinfo {year} {2012})\ Chap.~\bibinfo {chapter} {3}, pp.\
  \bibinfo {pages} {90--145},\ \bibinfo {edition} {2nd}\ ed.\BibitemShut
  {Stop}%
\bibitem [{\citenamefont {Boncelet}(2005)}]{Bon05}%
  \BibitemOpen
  \bibfield  {author} {\bibinfo {author} {\bibfnamefont {C.}~\bibnamefont
  {Boncelet}},\ }in\ \href@noop {} {\emph {\bibinfo {booktitle} {Handbook of
  Image and Video Processing}}},\ \bibinfo {editor} {edited by\ \bibinfo
  {editor} {\bibfnamefont {A.}~\bibnamefont {Bovik}}}\ (\bibinfo  {publisher}
  {Elsevier},\ \bibinfo {address} {Amsterdam, Netherlands},\ \bibinfo {year}
  {2005})\ pp.\ \bibinfo {pages} {397--409}\BibitemShut {NoStop}%
\bibitem [{\citenamefont {Riley}\ \emph
  {et~al.}(2006{\natexlab{b}})\citenamefont {Riley}, \citenamefont {Hobson},\
  and\ \citenamefont {J.}}]{RHB06_ch18}%
  \BibitemOpen
  \bibfield  {author} {\bibinfo {author} {\bibfnamefont {K.~F.}\ \bibnamefont
  {Riley}}, \bibinfo {author} {\bibfnamefont {M.~P.}\ \bibnamefont {Hobson}}, \
  and\ \bibinfo {author} {\bibfnamefont {B.~S.}\ \bibnamefont {J.}},\ }\enquote
  {\bibinfo {title} {Special functions},}\ \ (\bibinfo  {publisher} {Cambridge
  University Press},\ \bibinfo {address} {Cambridge, UK},\ \bibinfo {year}
  {2006})\ Chap.~\bibinfo {chapter} {18}, pp.\ \bibinfo {pages} {577--647},\
  \bibinfo {edition} {3rd}\ ed.\BibitemShut {Stop}%
\bibitem [{\citenamefont {Kish}\ \emph {et~al.}(2015)\citenamefont {Kish},
  \citenamefont {Granqvist}, \citenamefont {D{\'e}r},\ and\ \citenamefont
  {Kish}}]{KGDK2015}%
  \BibitemOpen
  \bibfield  {author} {\bibinfo {author} {\bibfnamefont {E.~A.}\ \bibnamefont
  {Kish}}, \bibinfo {author} {\bibfnamefont {C.-G.}\ \bibnamefont {Granqvist}},
  \bibinfo {author} {\bibfnamefont {A.}~\bibnamefont {D{\'e}r}}, \ and\
  \bibinfo {author} {\bibfnamefont {L.~B.}\ \bibnamefont {Kish}},\ }\href
  {\doibase 10.1007/s11571-015-9332-6} {\bibfield  {journal} {\bibinfo
  {journal} {Cogn. Neurodyn.}\ }\textbf {\bibinfo {volume} {9}},\ \bibinfo
  {pages} {459} (\bibinfo {year} {2015})}\BibitemShut {NoStop}%
\bibitem [{\citenamefont {Kai}\ \emph {et~al.}(1987)\citenamefont {Kai},
  \citenamefont {Higaki}, \citenamefont {Imasaki},\ and\ \citenamefont
  {Furukawa}}]{KSF87}%
  \BibitemOpen
  \bibfield  {author} {\bibinfo {author} {\bibfnamefont {S.}~\bibnamefont
  {Kai}}, \bibinfo {author} {\bibfnamefont {S.}~\bibnamefont {Higaki}},
  \bibinfo {author} {\bibfnamefont {M.}~\bibnamefont {Imasaki}}, \ and\
  \bibinfo {author} {\bibfnamefont {H.}~\bibnamefont {Furukawa}},\ }\href
  {\doibase 10.1103/PhysRevA.35.374} {\bibfield  {journal} {\bibinfo  {journal}
  {Phys. Rev. A}\ }\textbf {\bibinfo {volume} {35}},\ \bibinfo {pages} {374}
  (\bibinfo {year} {1987})}\BibitemShut {NoStop}%
\bibitem [{\citenamefont {Montgomery}\ \emph
  {et~al.}(2012{\natexlab{a}})\citenamefont {Montgomery}, \citenamefont
  {Peck},\ and\ \citenamefont {Vining}}]{MPV12_ch1}%
  \BibitemOpen
  \bibfield  {author} {\bibinfo {author} {\bibfnamefont {D.~C.}\ \bibnamefont
  {Montgomery}}, \bibinfo {author} {\bibfnamefont {E.~A.}\ \bibnamefont
  {Peck}}, \ and\ \bibinfo {author} {\bibfnamefont {G.~G.}\ \bibnamefont
  {Vining}},\ }\enquote {\bibinfo {title} {Introduction},}\ in\ \href@noop {}
  {\emph {\bibinfo {booktitle} {Introduction to Linear Regression Analysis}}},\
  \bibinfo {series and number} {Wiley Series in Probability and Statistics}\
  (\bibinfo  {publisher} {John Wiley \& Sons},\ \bibinfo {address} {Hoboken,
  NJ},\ \bibinfo {year} {2012})\ Chap.~\bibinfo {chapter} {1}, pp.\ \bibinfo
  {pages} {1--11},\ \bibinfo {edition} {5th}\ ed.\BibitemShut {Stop}%
\bibitem [{\citenamefont {Chatterjee}\ and\ \citenamefont
  {Simonoff}(2013{\natexlab{a}})}]{CS2013_ch2}%
  \BibitemOpen
  \bibfield  {author} {\bibinfo {author} {\bibfnamefont {S.}~\bibnamefont
  {Chatterjee}}\ and\ \bibinfo {author} {\bibfnamefont {J.~S.}\ \bibnamefont
  {Simonoff}},\ }\enquote {\bibinfo {title} {Model building},}\ in\ \href
  {\doibase 10.1002/9781118532843} {\emph {\bibinfo {booktitle} {Handbook of
  Regression Analysis}}}\ (\bibinfo  {publisher} {John Wiley \& Sons},\
  \bibinfo {address} {Hoboken, NJ},\ \bibinfo {year} {2013})\ Chap.~\bibinfo
  {chapter} {2}, pp.\ \bibinfo {pages} {23--52}\BibitemShut {NoStop}%
\bibitem [{\citenamefont {Chatterjee}\ and\ \citenamefont
  {Hadi}(2012)}]{CH12_ch1}%
  \BibitemOpen
  \bibfield  {author} {\bibinfo {author} {\bibfnamefont {S.}~\bibnamefont
  {Chatterjee}}\ and\ \bibinfo {author} {\bibfnamefont {A.~S.}\ \bibnamefont
  {Hadi}},\ }\enquote {\bibinfo {title} {Introduction},}\ in\ \href@noop {}
  {\emph {\bibinfo {booktitle} {Regression Analysis by Example}}},\ \bibinfo
  {series and number} {Wiley Series in Probablity and Statistics}\ (\bibinfo
  {publisher} {John Wiley \& Sons},\ \bibinfo {address} {Hoboken, NJ},\
  \bibinfo {year} {2012})\ Chap.~\bibinfo {chapter} {1}, pp.\ \bibinfo {pages}
  {1--24},\ \bibinfo {edition} {5th}\ ed.\BibitemShut {Stop}%
\bibitem [{\citenamefont {Montgomery}\ \emph
  {et~al.}(2012{\natexlab{b}})\citenamefont {Montgomery}, \citenamefont
  {Peck},\ and\ \citenamefont {Vining}}]{MPV12_ch2}%
  \BibitemOpen
  \bibfield  {author} {\bibinfo {author} {\bibfnamefont {D.~C.}\ \bibnamefont
  {Montgomery}}, \bibinfo {author} {\bibfnamefont {E.~A.}\ \bibnamefont
  {Peck}}, \ and\ \bibinfo {author} {\bibfnamefont {G.~G.}\ \bibnamefont
  {Vining}},\ }\enquote {\bibinfo {title} {Simple linear regression},}\ in\
  \href@noop {} {\emph {\bibinfo {booktitle} {Introduction to Linear Regression
  Analysis}}},\ \bibinfo {series and number} {Wiley Series in Probability and
  Statistics}\ (\bibinfo  {publisher} {John Wiley \& Sons},\ \bibinfo {address}
  {Hoboken, NJ},\ \bibinfo {year} {2012})\ Chap.~\bibinfo {chapter} {2}, pp.\
  \bibinfo {pages} {12--66},\ \bibinfo {edition} {5th}\ ed.\BibitemShut {Stop}%
\bibitem [{\citenamefont {Montgomery}\ \emph
  {et~al.}(2012{\natexlab{c}})\citenamefont {Montgomery}, \citenamefont
  {Peck},\ and\ \citenamefont {Vining}}]{MPV12_ch10}%
  \BibitemOpen
  \bibfield  {author} {\bibinfo {author} {\bibfnamefont {D.~C.}\ \bibnamefont
  {Montgomery}}, \bibinfo {author} {\bibfnamefont {E.~A.}\ \bibnamefont
  {Peck}}, \ and\ \bibinfo {author} {\bibfnamefont {G.~G.}\ \bibnamefont
  {Vining}},\ }\enquote {\bibinfo {title} {Variable selection and model
  building},}\ in\ \href@noop {} {\emph {\bibinfo {booktitle} {Introduction to
  Linear Regression Analysis}}},\ \bibinfo {series and number} {Wiley Series in
  Probability and Statistics}\ (\bibinfo  {publisher} {John Wiley \& Sons},\
  \bibinfo {address} {Hoboken, NJ},\ \bibinfo {year} {2012})\ Chap.~\bibinfo
  {chapter} {10}, pp.\ \bibinfo {pages} {327--371},\ \bibinfo {edition} {5th}\
  ed.\BibitemShut {Stop}%
\bibitem [{\citenamefont {Chatterjee}\ and\ \citenamefont
  {Simonoff}(2013{\natexlab{b}})}]{CS2013_ch1}%
  \BibitemOpen
  \bibfield  {author} {\bibinfo {author} {\bibfnamefont {S.}~\bibnamefont
  {Chatterjee}}\ and\ \bibinfo {author} {\bibfnamefont {J.~S.}\ \bibnamefont
  {Simonoff}},\ }\enquote {\bibinfo {title} {Multiple linear regression},}\ in\
  \href {\doibase 10.1002/9781118532843} {\emph {\bibinfo {booktitle} {Handbook
  of Regression Analysis}}}\ (\bibinfo  {publisher} {John Wiley \& Sons},\
  \bibinfo {address} {Hoboken, NJ},\ \bibinfo {year} {2013})\ Chap.~\bibinfo
  {chapter} {1}, pp.\ \bibinfo {pages} {3--22}\BibitemShut {NoStop}%
\bibitem [{\citenamefont {Palittapongarnpim}\ \emph
  {et~al.}(2017{\natexlab{b}})\citenamefont {Palittapongarnpim}, \citenamefont
  {Wittek},\ and\ \citenamefont {Sanders}}]{PWS17}%
  \BibitemOpen
  \bibfield  {author} {\bibinfo {author} {\bibfnamefont {P.}~\bibnamefont
  {Palittapongarnpim}}, \bibinfo {author} {\bibfnamefont {P.}~\bibnamefont
  {Wittek}}, \ and\ \bibinfo {author} {\bibfnamefont {B.~C.}\ \bibnamefont
  {Sanders}},\ }in\ \href {\doibase 10.1109/SMC.2017.8122618} {\emph {\bibinfo
  {booktitle} {2017 IEEE International Conference on Systems, Man, and
  Cybernetics (SMC)}}},\ \bibinfo {organization} {{IEEE}}\ (\bibinfo
  {publisher} {{IEEE} Systems, Man, and Cybernetics Society},\ \bibinfo
  {address} {Banff, AB},\ \bibinfo {year} {2017})\ pp.\ \bibinfo {pages}
  {294--299}\BibitemShut {NoStop}%
\bibitem [{\citenamefont {Jekel}(2017)}]{Jek17}%
  \BibitemOpen
  \bibfield  {author} {\bibinfo {author} {\bibfnamefont {C.}~\bibnamefont
  {Jekel}},\ }\href {https://github.com/cjekel/piecewise_linear_fit_py}
  {\enquote {\bibinfo {title} {Fitting a piecewise linear function to data},}\
  }\bibinfo {howpublished}
  {\url{https://jekel.me/2017/Fit-a-piecewise-linear-function-to-data/}}
  (\bibinfo {year} {2017})\BibitemShut {NoStop}%
\bibitem [{\citenamefont {Khaneja}\ \emph {et~al.}(2005)\citenamefont
  {Khaneja}, \citenamefont {Reiss}, \citenamefont {Kehlet}, \citenamefont
  {Schulte-Herbr\"{u}ggen},\ and\ \citenamefont {Glaser}}]{KRK+05}%
  \BibitemOpen
  \bibfield  {author} {\bibinfo {author} {\bibfnamefont {N.}~\bibnamefont
  {Khaneja}}, \bibinfo {author} {\bibfnamefont {T.}~\bibnamefont {Reiss}},
  \bibinfo {author} {\bibfnamefont {C.}~\bibnamefont {Kehlet}}, \bibinfo
  {author} {\bibfnamefont {T.}~\bibnamefont {Schulte-Herbr\"{u}ggen}}, \ and\
  \bibinfo {author} {\bibfnamefont {S.~J.}\ \bibnamefont {Glaser}},\
  }\href@noop {} {\bibfield  {journal} {\bibinfo  {journal} {J. Magn. Reson.}\
  }\textbf {\bibinfo {volume} {172}},\ \bibinfo {pages} {296} (\bibinfo {year}
  {2005})}\BibitemShut {NoStop}%
\end{thebibliography}%
\end{document}